\documentclass[11pt]{article}

\makeatletter
\@addtoreset{equation}{section}
\makeatother

\newcommand{\w}[1]{\\[0.#1cm]}

\def\eq#1{(\ref{#1})}

\def\ben{\begin{equation}}
\def\een{\end{equation}}

\def\eb{ {\bar\epsilon} }

\let\a=\alpha \let\b=\beta  \let\d=\delta\let\e=\epsilon
 \let\h=\eta  \let\i=\iota \let\k=\kappa
\let\l=\lambda \let\m=\mu \let\n=\nu 
\def\p{\partial}
\let\r=\rho \let\s=\sigma \let\t=\tau  \let\f=\phi
\let\c=\chi 

 \let\D=\Delta  \let\L=\Lambda
  \let\S=\Sigma  \let\F=\Phi 
\let\C=\Chi 
\let\la=\label

\def\nn{\nonumber} \def\bd{\begin{document}} \def\ed{\end{document}}

\def\ds{\documentstyle} \let\fr=\frac \let\bl=\bigl \let\br=\bigr

\let\Br=\Bigr \let\Bl=\Bigl
\let\na=\nabla
\let\pa=\partial \let\ov=\overline

\newcommand{\be}{\begin{equation}}
\newcommand{\ee}{\end{equation}}
\def\ba{\begin{array}}
\def\ea{\end{array}}

\def\ft#1#2{{\textstyle{{\scriptstyle #1}\over {\scriptstyle #2}}}}
\def\fft#1#2{{#1 \over #2}}
\def\del{\partial}
\def\vp{\varphi}
\def\sst#1{{\scriptscriptstyle #1}}
\def\oneone{\rlap 1\mkern4mu{\rm l}}
\def\td{\tilde}
\def\wtd{\widetilde}
\def\ie{\rm i.e.\ }
\def\dalemb#1#2{{\vbox{\hrule height .#2pt
     \hbox{\vrule width.#2pt height#1pt \kern#1pt\vrule width.#2pt}
     \hrule height.#2pt}}}
\def\square{\mathord{\dalemb{6.8}{7}\hbox{\hskip1pt}}}
\newcommand{\ho}[1]{$\, ^{#1}$}
\newcommand{\hoch}[1]{$\, ^{#1}$}
\newcommand{\bea}{\begin{eqnarray}}
\newcommand{\eea}{\end{eqnarray}}
\newcommand{\ra}{\rightarrow}
\newcommand{\lra}{\longrightarrow}
\newcommand{\Lra}{\Leftrightarrow}
\newcommand{\tr}{{\rm tr} }
\newcommand{\Tr}{{\rm Tr} }

\def\cA{{{\cal A}}}
\def\cF{{{\cal F}}}
\def\tV{\widetilde V}
\def\tW{\widetilde W}
\def\tH{\widetilde H}
\def\tE{\widetilde E}
\def\tF{\widetilde F}
\def\tA{\widetilde A}
\def\im{{{\rm i}}}
\def\tY{{{\wtd Y}}}
\def\ep{{\epsilon}}
\def\vep{{\varepsilon}}
\def\R{\rlap{\rm I}\mkern3mu{\rm R}}
\def\bD{{{\bar D}}}

\def\R{\rlap{\rm I}\mkern3mu{\rm R}}
\def\bD{{{\bar D}}}
\def\R{{{\Bbb R}}}
\def\C{{{\Bbb C}}}
\def\H{{{\Bbb H}}}
\def\CP{{{\Bbb C}{\Bbb P}}}
\def\RP{{{\Bbb R}{\Bbb P}}}
\def\Z{{{\Bbb Z}}}
\def\bA{{{\Bbb A}}}
\def\bB{{{\Bbb B}}}
\def\bC{{{\Bbb C}}}
\def\bR{{{\Bbb R}}}
\def\bD{{{\Bbb D}}}
\def\bE{{{\Bbb E}}}
\def\bZ{{{\Bbb Z}}}

\def\Re{{{\frak{Re}}}}
\def\Im{{{\frak{Im}}}}

\def\pb{{\bar\psi}}
\def\hp{{\widehat\psi}}
\def\eb{{\bar\epsilon}}
\def\af{e^{-\alpha\phi/2}}
\def\afp{e^{\alpha\phi/2}}

\def\a{\alpha}
\def\b{\beta}
\def\c{\gamma}\def\C{\Gamma}\def\cdt{\dot\gamma}
\def\d{\delta}\def\D{\Delta}\def\ddt{\dot\delta}
\def\e{\epsilon}\def\vare{\varepsilon}
\def\f{\phi}\def\F{\Phi}\def\vvf{\f}
\def\h{\eta}
\def\i{\iota}
\def\k{\kappa}
\def\l{\lambda}\def\L{\Lambda}
\def\m{\mu}
\def\n{\nu}
\def\r{\rho}
\def\s{\sigma}\def\S{\Sigma}
\def\t{\tau}

\def\ua{\underline{\alpha}}
\def\ub{\underline{\phantom{\alpha}}\!\!\!\beta}
\def\uc{\underline{\phantom{\alpha}}\!\!\!\gamma}
\def\ul{\underline{\lambda}}
\def\um{\underline{\mu}}
\def\ud{\underline\delta}
\def\ue{\underline\epsilon}
\def\uo{\underline\omega}\def\uO{\underline{\Omega}}
\def\una{\underline a}\def\unA{\underline A}
\def\unb{\underline b}\def\unB{\underline B}
\def\unc{\underline c}\def\unC{\underline C}
\def\und{\underline d}\def\unD{\underline D}
\def\une{\underline e}\def\unE{\underline E}
\def\unf{\underline{\phantom{e}}\!\!\!\! f}\def\unF{\underline F}
\def\ung{\underline g}\def\unG{\underline G}
\def\uni{\underline i}\def\unI{\underline I}
\def\unj{\underline j}\def\unJ{\underline J}
\def\unk{\underline k}\def\unK{\underline K}
\def\unl{\underline \ell}\def\unL{\underline L}
\def\unm{\underline m}\def\unM{\underline M}\def\unK{\underline K}
\def\unn{\underline n}\def\unN{\underline N}
\def\unp{\underline{\phantom{a}}\!\!\! p}\def\unP{\underline P}
\def\unq{\underline{\phantom{a}}\!\!\! q}
\def\unr{\underline{\phantom{a}}\!\!\! r}
\def\uns{\underline{\phantom{a}}\!\!\! s}
\def\ur{\underline r}
\def\us{\underline s}
\def\ut{\underline t}
\def\unQ{\underline{\phantom{A}}\!\!\!\! Q}
\def\uny{\underline{y}}
\def\unH{\underline{H}}
\def\unF{\underline{F}}\def\unT{\underline{T}}\def\unR{\underline{R}}
\def\unK{\underline{K}}
\def\wt{\widetilde}
\def\cV{{\mathcal V}}
\def\cP{{\mathcal P}}
\def\td{\tilde}
\def\wtd{\widetilde}
\def\hi{\hat i}
\def\hj{\hat j}
\def\hk{\hat k}
\def\hell{\hat\ell}
\def\hcV{\widehat\cV}
\def\hP{{\widehat P}}
\def\ef{e^{\varphi}}
\def\emf{e^{-\varphi}}
\def\es{e^{-\sigma/2}}
\def\eps{e^{\sigma/2}}
\def\hr{{\hat r}}
\def\hs{{\hat s}}
\def\hQ{{\widehat Q}}
\def\hC{{\hat C}}
\def\hL{{\hat L}}
\def\hL{{\hat L}}
\def\cL{{\mathcal L}}
\def\hf{{\widehat f}}
\def\hI{{\hat I}}
\def\hJ{{\hat J}}
\def\hK{{\hat K}}

\newcommand{\auth}{\large E. Bergshoeff\hoch{\dagger},
D.C. Jong\hoch{\ddagger} and E. Sezgin\hoch{\ddagger}}

\begin{document}
\begin{flushright}
\hfill{MIFP-05-12}\\
\hfill{UG-05-04}\\
\hfill{ \bf hep-th/yymmddd}\\
\today
\end{flushright}

\vspace{15pt}

\begin{center}


 {\Large \bf Noncompact Gaugings, Chiral Reduction }

 \medskip

 {\Large \bf and Dual Sigma Models in Supergravity}


\end{center}

\bigskip


\centerline{{\large \bf E. Bergshoeff $^\dagger$, D.C. Jong $^*$ and
E. Sezgin $^*$}}


\bigskip\bigskip

\begin{center}
{\it $^\dagger$ Institute for Theoretical Physics, Nijenborgh 4,
9747 AG Groningen, The Netherlands

\medskip

$^*$ George P. and Cynthia W. Mitchell Institute for Fundamental
Physics,Texas A\&M University, College Station, TX 77843-4242, U.S.A
}

\end{center}

\vspace{50pt}

\begin{center}

\underline{ABSTRACT}

\end{center}

\vspace{12pt}

We show that the half-maximal $SU(2)$ gauged supergravity with
topological mass term admits coupling of an arbitrary number of $n$
vector multiplets. The chiral circle reduction of the {\it ungauged}
theory in the dual 2-form formulation gives $N=(1,0)$ supergravity
in $6D$ coupled to $3p$ scalars that parametrize the coset
$SO(p,3)/SO(p)\times SO(3)$, a dilaton and $(p+3)$ axions with $p\le
n$. Demanding that $R$-symmetry gauging survives in $6D$ is shown to
put severe restrictions on the $7D$ model, in particular requiring
noncompact gaugings. We find that the $SO(2,2)$ and $SO(3,1)$ gauged
$7D$ supergravities give a $U(1)_R$, and the $SO(2,1)$ gauged $7D$
supergravity gives an $Sp(1)_R$ gauged chiral $6D$ supergravities
coupled to certain matter multiplets. In the $6D$ models obtained,
with or without gauging, we show that the scalar fields of the
matter sector parametrize the coset $SO(p+1,4)/SO(p+1)\times SO(4)$,
with the $(p+3)$ axions corresponding to its abelian isometries. In
the ungauged $6D$ models, upon dualizing the  axions to $4$-form
potentials, we obtain coupling of $p$ linear multiplets and one
special linear multiplet to chiral $6D$ supergravity.

 {\vfill\leftline{}\vfill \vskip
10pt \footnoterule

{\footnotesize \hoch{\dagger} Research
 supported in part by NSF Grant PHY-0314712 \vskip -12pt} \vskip 14pt }

\pagebreak \setcounter{page}{2}

\tableofcontents

\addtocontents{toc}{\protect\setcounter{tocdepth}{2}}

\newpage


\section{Introduction}


An impressively large number of string/M theory vacua admit a low
energy supergravity description in diverse dimensions in which,
however, the R-symmetry group is either trivial or a global
(ungauged) symmetry. There exists a far smaller class of vacua which
admit gauged supergravities, which, by definition are those in which
the R-symmetry is nontrivial and gauged by means of a vector field
inside or outside the supergravity multiplet, or a combination
thereof\footnote{When the gauge group is larger than the
$R$-symmetry group, they will still be referred to as gauged
supergravities as long as it contains the $R$-symmetry group.}.
These have played an important role in phenomena such as the AdS/CFT
as well as the domain-wall/quantum field theory correspondence. The
standard examples involve the maximally supersymmetric $SO(N)$
gauged supergravities but it has been shown that the noncompact
gauged supergravities, whose gauge group is a contraction and/or
analytic continuation of $SO(N)$, with less supersymmetries play a
role as well.

While a full classification of all possible gauged supergravities is
not available, it is natural to investigate whether all of the ones
that are known can be obtained from string/M theory. This is not a
simple problem, of course, when addressed in such a generality.
However, there is a class of gauged supergravities that are
especially interesting to explore, namely matter coupled gauged
minimal supergravities in six dimensions \cite{Nishino:1984gk}. One
immediate reason why these are interesting is that the requirement
of anomaly freedom turns out to be highly restrictive such that
there are very few models that satisfy these criteria
\cite{Randjbar-Daemi:1985wc,Salam:1985mi,Bergshoeff:1986hv,Avramis:2005qt,
Avramis:2005hc}, and moreover, so far it is not known if any of
these models can be embedded in string/M-theory.  We expect that if
and when such embeddings are discovered, they are likely to reveal
novel phenomena in string/M-theory.

Certain progress has already been made by Cvetic, Gibbons and Pope
\cite{Cvetic:2003xr} who showed that the $U(1)$ gauged minimal 6D
supergravity, while anomalous, it can nonetheless be obtained from
M-theory on $H_{2,2}\times S^1$ followed by a truncation. Here,
$H_{2,2}$ is a noncompact hyperboloidal 3-manifold that can be
embedded in a $(2,2)$ signature plane. The new phenomenon is that
the intermediary $7D$ theory obtained prior to the circle reduction
and chiral truncation must have a noncompact gauge group, in this
case $SO(2,2)$. Naive reduction attempts had not worked prior to
this observation.

The purpose of this paper is to start from a general gauged
supergravity theory in 7D \cite{Bergshoeff:1985mr}, and putting
aside its string/M-theory origin, as well as the issue of
anomaly-freedom for now, to look for the most general circle
reduction followed by chiral truncation with the criteria that a
\emph{gauged} 6D supergravity results. The generic {\it ungauged}
half-maximal supergravity coupled to $n$ vector multiplets
\cite{Bergshoeff:1985mr} has scalar fields whose interactions are
governed by the coset $SO(n,3)/SO(n)\times SO(3)$. The chiral
reduction of this model gives rise to an $N=(1,0)$ supergravity in
$6D$ coupled to $p\le n$ on-shell linear multiples, one on-shell
special linear multiplet and $(n-p)$ vector multiplets. The theory
in $7D$ is half-maximal and coupled to $n$ vector multiplets. Thus,
it has $(40+8n)_B + (40+8n)_F$ degrees of freedom, while the chiral
theory we end up with in $6D$ has $[16+ 4+ 4p + 4(n-p)]_B + [16+
4+4p +4(n-p)]_F$, that is to say $(20+4n)_B+(20+4n)_F$ degrees o
freedom, which, as expected, is half the degrees of freedom we start
with in $7D$.

An on-shell linear multiplet is the dual version of an off-shell
linear multiplet whose bosons consist of a 4-form potential and
three scalars, in which the 4-form potential is dualized to an
axionic scalar. What we call an on-shell {\it special linear
multiplet} has a bosonic sector that consists of three axionic
scalars that can be dualized to three $4$-form potentials and one
scalar. We show that the model can be reformulated as a $6D$
supergravity theory coupled to $(p+1)$ hypermultiplets whose scalar
fields parametrize the coset $SO(p+1,4)/SO(p+1)\times SO(4)$. This
will be studied both in the symmetric gauge as well as in the
Iwasawa gauge. The isometry group contains $(p+3)$ abelian
isometries which correspond to the $(n+3)$ axions that can be
dualized to the $p$ linear and one special linear multiplets
described above. This dualization is carried out here and the
results are presented in Appendix C.

All of this is similar to the hypermultiplet coupling to $N=2$
supergravity in $4D$ in which case the abelian isometries have a
dual description in terms of a suitable number of tensor multiplets,
which are the $4D$ version of our linear multiplets, and one
double-tensor multiplet, which is the analog of our special linear
multiplet (which we might view as triple-linear multiplet)
\cite{Anguelova:2002kd}.

Turning to the {\it gauged} half-maximal supergravities coupled to
Yang-Mills in $7D$, the allowed semi-simple gauge groups are of the
form

\be G_0\times H\subset SO(n,3)\ , \ee

with $G_0$ being one of the six groups listed in \eq{2a}. The
noncompact gauge groups among them are those which have up to $3$
compact or up to $3$ noncompact generators. The models of special
interest are those in which the chiral truncation of the $7D$ gauged
theory gives rise to an $R$-symmetry gauged theory in $6D$. Such
models are very difficult to obtain from higher dimensions, and
indeed, only few such models exist. The ones we find are $Sp(1)_R$
or $U(1)_R$ gauged and matter coupled $6D$ supergravities with
hidden $SO(p+1,4)/SO(p+1)\times SO(4)$ structure for $p=0,1$, and
one or no {\it external Maxwell multiplets} that do not participate
in the gauging of the $R$-symmetry. The content of these models will
be summarized below. Analogous models have been constructed directly
in $4D$ as $N=2$ gauged supergravity coupled to hypermultiplets in
which abelian isometries of the quaternionic Kahler scalar manifold
are dualized to tensors and they are known to arise in certain
compactifications of string theories
\cite{Theis:2003jj,Dall'Agata:2003yr}\footnote{In the context of
globally supersymmetric sigma models, the phenomenon of dualizing
abelian isometries of a hyperkahler manifold to obtain tensor
multiplets in $4D$, and linear multiplets in $6D$, was described
long ago in \cite{Howe:1985sb}. The $4D$ case was treated in more
detail in \cite{Theis:2003jj} as well.}.

As we shall see, the gauged chiral $6D$ supergravities arise from
half-maximal $7D$ supergravities with {\it noncompact} gaugings.
While noncompact gauging  is necessary, it is not sufficient for
obtaining $R$-symmetry gauging in $6D$. Indeed, of the five possible
noncompact gaugings listed in \eq{2a}, we shall find that the
$SL(3,R)$ gauged theory does not admit a chiral circle reduction to
a gauged $6D$ supergravity. In the case of $SO(2,1)^3$ gauged
theory, it turns out that a consistent chiral reduction with
surviving $6D$ gauge group $O(1,1)^3$ is possible but these do not
gauge the $R$-symmetry. The remaining three noncompact gauged
supergravities in $7D$, however, do give rise to the $R$-symmetry
gauged supergravities in $6D$ and we find the following three
models:

\begin{itemize}

\item {\bf The $SO(3,1)$ model:}

This model is obtained form the  $SO(3,1)$ gauged half-maximal $7D$
supergravity coupled to $3$ vector multiplets, with
$SO(3,3)/SO(3)\times SO(3)$ scalar sector. Its chiral reduction
gives a {\it $U(1)_R$ gauged supergravity coupled to a special
linear multiplet in $6D$}, referred to as model II in section 3.2.

\item {\bf The $SO(2,1)$ model:}

This model is obtained form the $SO(2,1)$ gauged half-maximal $7D$
supergravity coupled to a single vector multiplet, with
$SO(3,1)/SO(3)$ scalar sector. Its chiral reduction gives rise to an
{\it $Sp(1)_R$ gauged supergravity coupled to a special linear
multiplet in $6D$}, referred to as model IV in section 3.2.

\item  {\bf The $SO(2,2)$ model:}

This model is obtained form the $SO(2,2)$ gauged half-maximal $7D$
supergravity coupled to a $3$ vector multiplets, with
$SO(3,3)/SO(3)$ scalar sector. Its chiral reduction gives a {\it
$U(1)_R$ gauged theory coupled to an additional Maxwell multiplet, a
linear multiplet and a special linear multiplet in $6D$}, referred
to as model V in section 3.2.

\end{itemize}

The $SO(2,2)$ and $SO(3,1)$ models can be obtained from a reduction
of the $N=1, D=10$ supergravity on the noncompact hyperboloidal
3-manifolds  $H_{2,2}$ and $H_{3,1}$, respectively
\cite{Hull:1984vg,Hull:1988jw,Cvetic:2003xr}\footnote{These
reductions can straightforwardly be lifted to $D=11$. Note also that
the spaces $H_{p,q}$ can be constructed from embedding into a
$(p,q)$ signature plane. }. These models can also be obtained from
analytical continuation of an $SO(4)$ gauged $7D$ supergravity
\cite{Salam:1983fa} which, in turn, can be obtained from an $S^3$
compactification of Type IIA supergravity \cite{Cvetic:1999pu}, or a
limit of an $S^4$ reduction of $D=11$ supergravity which reduces to
a compactification on $S^3\times R$ \cite{Lu:1999bc}.

The specific gauged supergravities we have found are expected to
play a role in a string/M theoretic construction of an anomaly-free
matter coupled minimal gauged supergravity in $6D$. With regard to
various matter coupled $7D$ supergravities considered here, we note
that the heterotic string on $T^3$ gives rise to half-maximal $7D$
supergravity coupled to $19$ Maxwell multiplets, which, in turn, is
dual to M-theory on $K3$.

Finally, in this paper, we have also shown that, contrary to the
claims made in the literature \cite{Park:1988id}, we can dualize the
2-form potential to a 3-form potential in the gauged $7D$
supergravity even in the presence of couplings to an arbitrary
number of vector multiplets (see Appendix A).

This paper is organized as follows. In section 2, we recall the
(gauged) half-maximal supergravity couple to $n$ vector multiplets
\cite{Bergshoeff:1985mr}. In particular, we list the possible
non-compact gaugings in this theory. In section 3, we determine the
conditions that must be satisfied by the requirement of chiral
supersymmetry in $6D$, both, for the gauged and ungauged $7D$
theory. We then solve these conditions, and determine  the field
content, the $6D$ supermultiplet structure. In section 4, we obtain
the $6D$ supergravity for the fields that survive the chiral
reduction, and their supersymmetry transformation rules. In section
5, we exhibit the hidden quaternionic Kahler coset structure that
given the couplings of the matter multiplets in $6D$ by an extensive
use of the Iwasawa decompositions. This is not surprising for the
ungauged and bosonic sector. Here, we show the result for the gauged
and ungauged models, and including the fermionic sectors well. A
brief summary and comments are given in section 7.

In Appendix A, we give our result for the dual formulation of the
gauged and matter coupled $7D$ supergravity, in which the 2-form
potential is dualized to a 3-form potential. A useful relation
between the $SL(4,R)/SO(4)$ parametrization used in
\cite{Salam:1983fa, Cvetic:2003xr} and the $SO(3,3)/SO(3)\times
SO(3)$ parametrization used in \cite{Bergshoeff:1985mr} and in this
paper. In Appendix C, considering the ungauged $6D$ models, we
dualize the axionic scalar field to $4$-form potentials, thereby
obtaining a coupling of a an arbitrary number of linear multiplets
to a single special linear multiplet. Appendix D,contains some
useful formula on the Iwasawa decomposition of $SO(p,q)$ that is
used in showing  the hidden quaternionic Kahler coset structure in
six dimensional model. that we obtain by chiral reduction.


\section{The Gauged $7D$ Model with Matter Couplings}


Half-maximal supergravity in $D=7$ coupled to $n$
vector multiplets has the field content
 \be
 \left(e_\mu{}^m, B_{\mu\nu}, \sigma, A_\mu^I, \phi^\a, \psi_\mu,
 \chi,\lambda^r\right)\ ,
 \ee

where the fermions $\psi_\mu, \chi,\lambda^r$ are symplectic
Majorana and they all carry $Sp(1)$ doublet indices which have been
suppressed. The $3n$ scalars $\phi^\a (\a=1,2,..,3n)$ parametrize
the coset
 \be
 {SO(n,3)\over SO(n)\times SO(3)}\ .
 \label{2coset}
 \ee

The gauge fermions $\lambda^r\, (r=1,...,n)$ transform in the vector
representation of $SO(n)$, while the vector fields $A_\mu^I
\,(I=1,...,n+3)$ transform in the vector representation of
$SO(n,3)$. The 2-form potential $B_{\mu\nu}$ and the dilaton
$\sigma$ are real.

It is useful to define a  few ingredients associated with the scalar
coset manifold as they arise in the Lagrangian. We first introducing the coset
 representative
 \be
 L=(L_I{}^i, L_I{}^r)\ ,\qquad I=1,...,n+3,\quad i=1,2,3\ ,\qquad
 r=1,...,n\ ,
 \label{2L}
 \ee

 which forms an $(n+3)\times(n+3)$ matrix that obeys
 the relation
 \be
 -L_I{}^iL_J^i + L_I{}^r L_J{}^r=\eta_{IJ}\ ,
 \label{br}
 \ee

where $\eta_{IJ}= {\rm diag} (---++...+)$. The contraction of the
$SO(n)$ and $SO(3)$ indices is with the Kronecker deltas
$\delta_{rs}$ and $\delta_{ij}$ while the raising and lowering of
the $SO(n,3)$ indices will be with the $SO(n,3)$ invariant metric
$\eta_{IJ}$. Given that the $SO(3)$ indices are raised and lowered
by the Kronecker delta, it follows that, in our conventions,

 $$
  L_I^i=L_{Ii}\ ,\qquad L_I^i L^I_j= -\delta^i_j\ ,\qquad
  L_I^i L^{Ij}= -\delta^{ij}\ .
  $$

Note also that the inverse coset representative $L^{-1}$ is given by
$L^{-1}=\left(L^I{}_i, L^I{}_r\right)$ where
$L^I{}_i=\eta^{IJ}\,L_{Ji}$ and $L^I{}_r=\eta^{IJ}\,L_{Jr}$. In the
gauged matter coupled theory of \cite{Bergshoeff:1985mr}, a key
building block is the gauged Maurer-Cartan form
 \bea
 P_\mu^{ir} &=& L^{Ir}\left(\p_\mu\delta_I^K + f_{IJ}{}^K A_\mu^J \right)
 L_K^i\ ,\nn\w2
 Q_\mu^{ij} &=& L^{Ij}\left(\p_\mu\delta_I^K + f_{IJ}{}^K A_\mu^J \right)
 L_K^i\ ,\nn\w2
 Q_\mu^{rs} &=& L^{Ir}\left(\p_\mu\delta_I^K + f_{IJ}{}^K A_\mu^J \right)
 L_K^{s}\ ,
 \label{2mc}
 \eea

where $f_{IJ}{}^K$ are the structure constants of the not
necessarily simple group $K \subset SO(n,3)$ of dimension $n+3$, and
the gauge coupling constants are absorbed into their definition of
the structure constants. The $K$-invariance of the theory requires
that the adjoint representation of $K$ leaves $\eta_{IJ}$ invariant:
 \be
 f_{IK}{}^L\,\eta_{LJ}+ f_{JK}{}^L\,\eta_{LI}=0\ .
 \label{ic}
 \ee

It follows that for each simple subgroup of $K$, the corresponding
part of $\eta_{IJ}$ must be a multiple of its Cartan-Killing metric.
Since $\eta_{IJ}$ contains an arbitrary number of positive entries,
$K$ can be an arbitrarily large compact group. On the other hand, as
$\eta_{IJ}$ has only three negative entries, $K$ can have $3$ or
less compact generators, or $3$ or less noncompact
generators\footnote{This is similar to the reasoning in
\cite{deRoo:1985jh} where the gauging of $N=4, D=4$ supergravity coupled
to $n$ vector multiplets is considered. In this case, the relevant
$\eta$ is the $SO(n,6)$ invariant tensor and the resulting
noncompact simple gauge groups have been listed in
\cite{deRoo:1985jh}.} . Thus, the allowed semi-simple gauge groups
are of the form $G_0 \times H \subset SO(n,3)$ where $G_0$ is one of
the following
 \bea
 (I) &&\ SO(3)\nn\\
 (II) &&\ SO(3,1)\nn\\
 (III) &&\ SL(3,R)\nn\\
 (IV) &&\ SO(2,1)\nn\\
 (V) &&\ SO(2,1)\times SO(2,1)\nn\\
 (VI) &&\ SO(2,1)\times SO(2,1)\times SO(2,1)
 \label{2a}
 \eea
and $H$ is a semi-simple compact Lie group with ${\rm dim}\, H\le
(n+3-{\rm dim}\,G_0 )$. Of these cases, only (I) with $H=SO(3)$
corresponding to SO(4) gauged supergravity, (II) and (V) are known to
have a ten- or eleven-dimensional origin. Though the cases (III)--(VI) are
not mentioned explicitly in \cite{Bergshoeff:1985mr}, the Lagrangian
provided there is valid for all the cases listed above.

The Lagrangian of \cite{Bergshoeff:1985mr}, up to quartic fermion
terms, is given by \footnote{We follow the conventions of
\cite{Bergshoeff:1985mr}. In particular,  $\eta_{\mu\nu}={\rm diag}
(-++ \cdots +)$, the spinors are symplectic Majorana, $C^T=C$ and
$(\c^\mu C)^T=-\c^\mu C$. Thus, ${\bar\psi}\c^{\nu_1\cdots\nu_n}\lambda=
(-1)^n{\bar\psi}\c^{\nu_n\cdots\nu_1}\lambda$, where the $Sp(1)$ doublet indices
are contracted and suppressed. Here we also use $X^A{}_B=\frac{1}{\sqrt
2} (\sigma^i)^A{}_B X^i$, and further conventions are: $X^A=\e^{AB}X_B,\ X_A=X^B\e_{BA},
\ \e^{AB}\e_{BC}=-\d^A_C,\ {\bar\psi}\lambda =\psi^A\lambda_A,\
{\bar\psi}\sigma^i\lambda={\bar\psi}^A(\sigma^i)_A{}^B\e_B$.}.
 \bea
 {\mathcal L} &=& {\mathcal L}_B + {\cal L}_F
 \label{2t}\w2
 e^{-1}{\cal L}_B &=& \ft12 R -\ft14 e^\sigma
 \left( F_{\mu\nu}^iF^{\mu\nu}_i+F_{\mu\nu}^rF^{\mu\nu}_r\right)
 -\ft1{12}e^{2\sigma} G_{\mu\nu\rho}G^{\mu\nu\rho}
 \nn\w2
 && -\ft58 \p_\mu\sigma \p^\mu \sigma - \ft12 P_\mu^{ir}
 P^\mu_{ir}-\ft14 e^{-\sigma} \left(C^{ir} C_{ir}-\ft19 C^2\right)\ ,
 \label{2b1}\w4
 e^{-1}{\cal L}_F &=&-\ft{i}2
 {\bar\psi}_\mu\c^{\mu\nu\rho}D_\nu\psi_\rho -\ft{5i}2
 {\bar\chi}\c^\mu D_\mu\chi -\ft{i}2 {\bar\lambda}^r
 \c^\mu D_\mu\lambda_r -\ft{5i}4 {\bar\chi}
 \c^\mu\c^\nu\psi_\mu \p_\nu\sigma
 \nn\w2
  &&-\ft12 {\bar\lambda}^r \sigma^i\c^\mu\c^\nu\psi_\mu P_{\nu ri}
  +\ft{i}{24\sqrt 2} e^\sigma G_{\mu\nu\rho} X^{\mu\nu\rho}
 +\ft18 e^{\sigma/2} F_{\mu\nu}^i X_i^{\mu\nu}
 -\ft{i}{4} e^{\sigma/2} F_{\mu\nu}^r X_r^{\mu\nu}
 \nn\w2
 &&-\ft{i\sqrt 2}{24} e^{-\sigma/2} C\left(
 {\bar\psi}_\mu\c^{\mu\nu}\psi_\nu
 +2{\bar\psi}_\mu\c^\mu\chi+3{\bar\chi}\chi-{\bar\lambda}^r\lambda_r
 \right)
 \nn\w2
 &&+\ft1{2\sqrt 2} e^{-\sigma/2} C_{ir}
 \left( {\bar\psi}_\mu \sigma^i\c^\mu\lambda^r
 -2{\bar\chi}\sigma^i\lambda^r\right)+\ft12
 e^{-\sigma/2}C_{rsi} {\bar\lambda}^r\sigma^i\lambda^s\ ,
 \label{2f}
 \eea
where the fermionic bilinears are defined as
 \bea
 X^{\mu\nu\rho}&=& {\bar\psi}^\lambda\c_{[\lambda}
 \c^{\mu\nu\rho}\c_{\tau]} \psi^\tau
 +
 4{\bar\psi}_\lambda\c^{\mu\nu\rho}\c^\lambda\chi
 -3{\bar\chi}\c^{\mu\nu\rho}\chi
 +{\bar\lambda}^r\c^{\mu\nu\rho}\lambda_r\ ,
 \nn\w2
 X_i^{\mu\nu}&=& {\bar\psi}^\lambda \sigma^i \c_{[\lambda}
 \c^{\mu\nu}\c_{\tau]} \psi^\tau
 -2{\bar\psi}_\lambda\sigma^i\c^{\mu\nu}\c^\lambda\chi
 +3{\bar\chi}\sigma^i\c^{\mu\nu}\chi
 -{\bar\lambda}^r\sigma^i\c^{\mu\nu}\lambda_r\ ,
 \nn\w2
 X_r^{\mu\nu}&=& {\bar\psi}_\lambda
 \c^{\mu\nu}\c^\lambda\lambda^r
 +2{\bar \chi}\c^{\mu\nu}\lambda^r\ .
 \eea

The field strengths and the covariant derivatives are defined as
 \bea
 && G_{\mu\nu\rho}= 3\p_{[\mu} B_{\nu\rho]}-\ft3{\sqrt 2}
 \omega^0_{\mu\nu\rho}\ ,\qquad \omega^0_{\mu\nu\rho} =
 F^I_{[\mu\nu} A_{\rho]}^J\eta_{IJ}-\ft13 f_{IJ}{}^K
 A_\mu^I A_\nu^J A_{\rho K} \ ,
 \nn\w2
 && F_{\mu\nu}^I = 2\p_{[\mu} A_{\nu]}^I +f_{JK}{}^I
 A_\mu^J A_\nu^K\ ,\qquad F^i_{\mu\nu}=F_{\mu\nu}^I L_I^i\ ,
 \qquad F^r_{\mu\nu}=F_{\mu\nu}^I L_I^r\ ,
 \w2
 &&D_\mu =
 \p_\mu+\frac14\omega_\mu{}^{ab}\c_{ab}+\frac{1}{2\sqrt
 2}Q_\mu^i\,\sigma^i\ ,\quad\quad Q_\mu^i=\frac{i}{\sqrt 2}~\epsilon^{ijk}
Q_{\mu
 jk}\ ,\label{cd}
 \eea
and the $C$-functions are given by \cite{Bergshoeff:1985mr}
 \bea
 &&C=-\ft{1}{\sqrt 2} f_{IJ}{}^K L^I_i L^J_j L_{K k}\,\epsilon^{ijk}\ ,
 \nn\w2
 &&C_{ir}=\ft1{\sqrt 2} f_{IJ}{}^K L^I_j L^J_k L_{K r}\,\epsilon^{ijk}\ ,
 \nn\w2
 &&C_{rsi}= f_{IJ}{}^K L^I_r L^J_s L_{K i}\ .
 \label{cf}
 \eea

 The local supersymmetry transformation rules read
 \cite{Bergshoeff:1985mr}
 \bea
 \delta e_\mu{}^m &=& i\eb \c^m\psi_\mu\ ,
 \nn\w2
 \delta\psi_\mu &=& 2D_\mu\epsilon
 -\ft1{60\sqrt 2}e^\sigma
 G_{\rho\sigma\tau}\left(\c_\mu\c^{\rho\sigma\tau}
 +5\c^{\rho\sigma\tau}\c_\mu \right)\epsilon
 \nn\w2
 &&-\ft{i}{20} e^{\sigma/2} F_{\rho\sigma}^i\, \sigma^i
 \left(3\c_\mu\c^{\rho\sigma}-5\c^{\rho\sigma}\c_\mu\right)
 \epsilon -\ft{\sqrt 2}{30} e^{-\sigma/2}C\c_\mu\epsilon\ ,
 \nn\w3
 \delta \chi&=&-\ft12\c^\mu\p_\mu\sigma\epsilon
 -\ft{i}{10} e^{\sigma/2}F_{\mu\nu}^i\,\sigma^i \c^{\mu\nu}\epsilon
 -\ft1{15\sqrt 2}
 e^\sigma G_{\mu\nu\rho}\c^{\mu\nu\rho}\epsilon+\ft{\sqrt
 2}{30}e^{-\sigma/2}C\epsilon\ ,
 \nn\w2
 \delta B_{\mu\nu}&=& i{\sqrt 2} e^{-\sigma}
 \left( \eb\c_{[\mu}\psi_{\nu]}+\eb\c_{\mu\nu}\chi\right)
 -{\sqrt 2}A_{[\mu}^I\delta A_{\nu]}^J\eta_{IJ}\ ,
 \nn\w2
 \delta \sigma &=& -2i\eb\chi\ ,
 \nn\w2
 \delta A_\mu^I &=& -e^{-\sigma/2}\left(\eb\sigma^i\psi_\mu
 +\eb\sigma^i\c_\mu\chi\right) L^I_i
 +ie^{-\sigma/2} \eb\c_\mu\lambda^r L^I_r\ ,
 \nn\w2
 \delta L_I^r &=& \eb\sigma^i \lambda^r L_I^i\ ,\quad\quad
 \delta L_I^i = \eb\sigma^i \lambda_r L_I^r\ ,
 \label{2s}\w2
 \delta \lambda^r &=& -\ft12 e^{\sigma/2}F_{\mu\nu}^r
 \c^{\mu\nu}\epsilon +i\c^\mu P_\mu^{ir}\sigma^i\epsilon
 -\ft{i}{\sqrt 2} e^{-\sigma/2} C^{ir} \sigma^i \epsilon\ .
 \nn
 \eea

For  purposes of the next section, we exhibit the gauge field dependent
part of the gauged Maurer-Cartan forms:
 \bea
 P_\mu^{ir} &=& P_\mu^{ir (0)}
 -\ft1{2\sqrt 2}\,\epsilon^{ijk}\,C^{jr} A_\mu^k -C^{irs}A_\mu^s\ ,
 \nn\w2
 Q_\mu^{ij} &=& Q_\mu^{ij(0)}+\ft1{3\sqrt
 2}\,\epsilon^{ijk}\,C\,A_\mu^k - \ft1{2\sqrt 2}\,\epsilon^{ijk}\,C^{kr}\,A_\mu^r\
 ,
 \label{3mc}
 \eea

where the zero superscript indicates the gauge field independent
parts.


\section{Chiral Reduction on a Circle}


\subsection{Reduction Conditions}


Here we shall consider all the $7D$ quantities of the previous section
such as fields, world and Lorentz indices to be hatted, and the corresponding
$6D$ quantities to be unhatted ones. We parametrize the $7D$ metric as
 \be
 d{\widehat s}^2=e^{2\a\phi}\,ds^2+e^{2\b\phi} (dy-{\cal A})^2\ .
 \ee

In order to obtain the canonical Hilbert-Einstein term in $D=6$, we
choose
 \be
 \alpha=-\frac1{2 \sqrt{10}}\ ,\qquad \beta= -4\alpha\ .
 \ee

We shall work with the natural vielbein basis
 \be
 {\hat e}^a=e^{\alpha\phi} e^a\ ,\quad\quad
 {\hat e}^7=e^{\beta\phi}(dy-{\cal A})\ .
 \ee

Next, we analyze the constraints that come from the requirement of
circle reduction followed by chiral truncation retaining $N=(1,0)$
supersymmetry. Let us first set to zero the $7D$ gauge coupling
constant and deduce the consistent chiral truncation
conditions. At the end of the section we shall then re-introduce the
 coupling constant and determine the additional constraints that
need to be satisfied.

The gravitino field in seven dimensions splits into a left handed
and a right handed gravitino in six dimensions upon  reduction in a
compact direction. Chiral truncation means that we set one of them
to zero, say,
 \be
 {\widehat \psi}_{a-}=0\ .
 \label{4g}
 \ee

This condition, used in the supersymmetry variation of the vielbein,
gravitino and the field ${\widehat \chi}$ readily gives the
following further conditions
\bea
 &&{\widehat F}_{ab}^{\hI}\,L_{\hI}{}^i=0\ ,
 \label{fc}
 \w2
 &&{\cal A}_a = 0\ ,\qquad {\widehat G}_{ab7}=0\ ,
 \qquad {\widehat \psi}_{7+}=0\ ,\qquad {\widehat \chi}_+=0\ .
 \label{oc}
 \eea

To see how we can satisfy the condition \eq{fc}, it is useful to
consider an explicit realization of the $SO(n,3)/SO(n)\times SO(3)$
coset representative. A convenient such parametrization is given by
 \be
 {\widehat L} = \left(\ba{ccc} {1+\f^t\f\over 1-\f^t\f} && {2\over 1-\f^t\f} \f^t\\
 &\\ \f {2\over 1-\f^t\f} && 1 + \f {2\over 1-\f^t\f} \f^t \ea
 \right)
 \label{4cr}
 \ee
\\
where $\f$ is a $n\times 3$ matrix $\phi_{{\hat r}i}$. Note that
this is symmetric, and as such, we shall refer to this as the coset
representative in the {\it symmetric gauge}. Now, we observe that to
satisfy \eq{fc}, we can split the index
 \be
 {\hI}=\{I,I'\}\ ,\qquad I=1,...,p+3\ ,\qquad I'=p+4,...,n+3\ ,
 \label{p}
 \ee
and set
 \be
 {\widehat A}_a^I=0\ ,\qquad L_{I'}{}^i=0\ .
 \label{L1}
 \ee

Note that $0\le p \le n$, and in particular, for $p=n$, all vector
fields $A_\mu^I, I=1,...,n+3$ vanish (i.e. there are no $A_\mu^{r'}$
fields) while all the coset scalars $\phi^{{\hat r}i}$ are
nonvanishing \footnote{ Note also that  for $p=0$, all coset scalars
$\phi^{{\hat r}i}$ vanish while $n$ vector field $A_\mu^{r'}$
survive.}. For $p<n$, however, as we shall see below, $(n-p)$ vector
fields survive, and these, in turn, will play a role in obtaining a
gauged supergravity in $6D$.

The second condition in \eq{L1} amounts to setting $\phi_{r'i}=0$ and
consequently, introducing the notation
 \be
 {\widehat r}=\{r,r'\}\ ,\qquad r=1,...,p\ ,\qquad r'=p+1,...,n\ ,
 \ee
we have
 \be
 L_I{}^{r'}=0\ ,\qquad L_{I'}{}^r=0\ ,\qquad
 L_{I'}{}^{r'}=\delta_{I'}{}^{r'}\ .
 \label{L2}
 \ee

Thus the surviving scalar fields are
 \be
 \left( {\widehat L}_I{}^i,{\widehat L}_I{}^r\right)\equiv
 \left( L_I{}^i,L_I{}^r\right)\ , \qquad I=1,...,p+3\ ,
 \ \ i=1,2,3\ ,\ \ r=1,...,p\ .
 \ee

This is the coset representative of $SO(p,3)/SO(p)\times SO(3)$. From the
supersymmetric variations of the vanishing coset representatives $(L_{I'}{}^i,
L_I{}^{r'},L_{I'}{}^r)$, on the other hand, we find that

 \be
 {\widehat\lambda}_{+}^r =0\ ,\quad {\widehat\lambda}_{-}^{r'}=0\ .
 \ee

Using  these results in the supersymmetry variation of ${\widehat
A}_7^{I'}$, in turn, immediately gives
 \be
 {\widehat A}_7^{I'}= 0\  .
 \ee

Next, defining
\be
    {\widehat B} = B_{\mu\nu}~dx^{\mu} \wedge dx^{\nu}
    +B_{\mu}~dx^{\mu} \wedge dy\ ,
 \ee

the already found conditions $\widehat{G}_{ab7}={\cal A}_a= {\widehat
A}_a^I=0$ imply that
 \be
 B_{\mu}= 0\ .
 \ee

In summary, the surviving bosonic fields are
 \be
 \left(g_{\mu\nu},\ \phi,\ B_{\mu\nu},\ {\widehat
 \sigma},\ \phi_{ir},\ {\widehat A}_7^I\ ,\ A_\mu^{I'}\right)\ ,
 \ee

and the surviving fermionic fields are
 \be
 \left( {\widehat \psi}_{\mu +},\ {\widehat \psi}_{7-},\ {\widehat \chi}_{-},
 \ {\widehat\lambda}^r_{-},\ {\widehat\lambda}^{r'}_{+}\right)\ .
 \ee

We will show in the next section that suitable combinations of these
fields (see Eq. \eq{defs8}) form the following supermultiplets:
 \bea
&& (g_{\mu\nu}, B_{\mu\nu}, \sigma,
     {\psi}_{\mu},\chi)\ , \ \
     (A_{\mu}^{I'}, \lambda^{r'})\ ,\ \
     (\phi_{ir}, \Phi^{I},\varphi,\lambda^r,\psi)\ ,\label{sm}
     \w2
     && I=1,..., p+3\ ,\quad I'=p+4,...,n+3\ ,\nn\\
     &&
     r=1,...,p\ ,\quad r'=p+1,...,n\ ,\quad i=1,2,3\ .\nn
 \eea
The last multiplet represents a fusion of $p$ linear multiplets and
one special linear multiplet, as explained in the introduction. In
particular, the $(p+3)$ axionic scalars $\Phi^I$ can be dualized to
$4$-form potentials. Further truncations are possible. Setting
$\phi_{ir}=0$ gives one special linear multiplet with fields
$(\Phi^i,\varphi,\psi)$ while setting $\Phi^I=0$ eliminates all the
(special) linear multiplets.\\

\centerline{\it Extra Conditions due to Gauging}
\bigskip

Extra conditions emerge upon turning on the $7D$ gauge coupling
constants. They arise from the requirement that the gauge coupling
constant dependent terms in the supersymmetry variations of $(
{\widehat\psi}_{a-},{\widehat\psi}_{7+},{\widehat\lambda}^r_+,
{\widehat\lambda}^{r'}_- )$ vanish. These conditions are
 \bea
 && C=0\ ,\qquad \qquad \ C^{ir}=0\ ,
 \label{c1}
 \nn\w2
 && C^{irs}\Phi^s=0\ ,\qquad C^{ir's'} A_\mu^{s'}=0\ ,
 \label{c2}
 \eea

where $\Phi^r=\Phi^I L_I^r$ and $A_\mu^{s'}=A_\mu^{I'}L_{I'}^{r'}$.
More explicitly, these conditions take the form
 \bea
 && \hf_{IJK}\,L^I_iL^J_jL^K_k=0\ ,\qquad \hf_{IJK}\,L^I_iL^J_jL^K_r=0\ ,
 \label{c3}\w2
 && \hf_{IJK}\,L^I_iL^J_r \Phi^K=0\ ,
 \qquad \hf_{Ir's'}\,L^I_i\,A_\mu^{s'}=0\ .
 \label{c4}
 \eea

Solving these conditions, while keeping all $A_\mu^{r'}$ and $\Phi^I$,
results in a chiral gauged supergravity theory with the multiplets shown
in \eq{sm} and gauge group $K'\subset SO(n,3)$ with structure constants
${\widehat f}_{r's't'}$. The scalars $\Phi^I$ transform in a $(p+3)$
dimensional representation of $K'$, and there are $3p$ scalars which
parametrize the coset $SO(p,3)/SO(p)\times SO(3)$. The nature of the
$R$-symmetry gauge group can be read off from
 \be
  D_\mu\epsilon= D_\mu^{(0)}\e+
  \ft1{2\sqrt 2}\sigma^i\,C^{ir'}\,A_\mu^{r'}\e\ .
  \label{r}
 \ee

Note that the $6D$ model is $R$-symmetry gauged provided that
$C^{ir'}$ does not vanish upon setting all scalars to zero.
Moreover, an abelian $R$-symmetry group can arise when
$\hf_{r's't'}$ vanishes with  $C^{kr'}\ne 0$. Next, we show how to
solve the conditions \eq{c3} and \eq{c4}.


\subsection{Solution to the Reduction Conditions}


The conditions \eq{c3} and \eq{c4} can be solved by setting
 \be
 \hf_{IJ}{}^K=0\ ,\qquad \hf_{I'J'}{}^K=0\ .
 \label{c11}
 \ee

Moreover, the structure constants of the $7D$ gauge group $G_0\times
H\subset SO(n,3)$ must satisfy the condition \eq{ic}:
 \be
 \hf_{\hI\hK}{}^{\hL}\,\eta_{\hL\hJ}
 + \hf_{\hJ\hK}{}^{\hL}\,\eta_{\hL\hI}=0\ .
 \label{ic2}
 \ee

Given the $7D$ gauge groups listed in \eq{2a}, we now check case by
case when and how these conditions can be satisfied. To begin with,
we observe that given the $G_0\times H \subset SO(n,3)$ gauged
supergravity theory, the $H$ sector can always be carried over to $6D$
dimension to give the corresponding Yang-Mills sector whose
$H$-valued gauge fields do not participate in a possible $R$-symmetry
gauging. Therefore, we shall consider the $G_0$ part of the $7D$
gauge group in what follows.


\subsubsection*{(I)\ \ $SO(3)$}


In this model, the $7D$ gauge group is $SO(3)$ with structure
constants
 \be
 \hf_{\hI\hJ\hK}=\left( g \,\e_{IJK}, 0 \right)\ .
 \ee

To satisfy \eq{c11}, we must set $g=0$. Thus, we see that {\it a
chiral truncation to a gauged $6D$ theory is not possible
 in this case}.


\subsubsection*{(II)\ \ $SO(3,1)$}


The smallest $7D$ scalar manifold that can accommodate this gauging
is  $SO(3,3)/SO(3)\times SO(3)$. In the $7D$ theory, the gauge group
$G_0=SO(3,1)$ can be embedded in $SO(3,3)$ as follows. Denoting the
$SO(3,3)$ generators by $T_{AB}=\left(T_{ij}, T_{rs},
T_{ir}\right)$, we can embed $SO(3,1)$ by choosing the generators
$(T_{rs}, T_{3r})$ which obey the commutation rules of the $SO(3,1)$
algebra. These generators can be relabeled as
 \be
 \left( T_{34},
T_{35},
T_{36},T_{45},T_{56},T_{64}\right)=\left(T_1,T_2,T_3,T_4,T_5,T_6\right)\equiv
 \left(T_I,T_{I'}\right)\ ,
 \ee

with $I=1,2,3$ and $I'=4,5,6$. The algebra of these generators is given
by
\be
 [T_{I},T_{J}]=f_{IJ}{}^{K'}\,T_{K'}\ ,
 \qquad
 [T_{I'},T_{J}]=f_{I'J}{}^{K}\,T_K\ ,
 \qquad
 [T_{I'},T_{J'}]=f_{I'J'}{}^{K'}\,T_{K'}\ .
 \label{h}
 \ee

Thus, the conditions \eq{c11} are satisfied. Furthermore, the
Cartan-Killing metric associated with this algebra is $(+++---)$ and
it satisfies the condition \eq{ic2}. In this case, all the coset
scalars are vanishing and the surviving matter scalar fields are
$(\Phi^i, \varphi)$ which are the bosonic fields of a special linear
multiplet. This sector will be shown to be described by the
quaternionic Kahler coset $SO(4,1)/SO(4)$ in section 5. We thus
obtain an {\it $Sp(1,R)$ gauged supergravity in $6D$ coupled to a
single hypermultiplet}. In summary, we have the following chain of
chiral circle reduction and hidden symmetry in this case:
\bea
 {SO(3,3)\over SO(3)\times SO(3)}\ \ \lra & (\Phi^i,\varphi) \  \lra  & \ \
 {SO(4,1)\over SO(4)}
 \label{h1}
\eea

Note that the  $7D$ theory we start with has $64_B+64_F$ physical
degrees of freedom, while the resulting $6D$ theory has $24_B+24_F$
physical degrees of freedom.


\subsubsection*{(III)\ \ $SL(3,R)$}


The minimal $7D$ scalar manifold to accommodate this gauging is
$SO(5,3)/SO(5)\times SO(3)$. In the $7D$ theory, the gauge group is
$SL(3,R)$, which has $3$ compact and $5$ noncompact generators. The
condition \eq{ic2} can be satisfied with $\eta={\rm diag}(---+++++)$ by
making a particular choice of the generators of $SL(3,R)$ such as
 \be
  (i\lambda_2,i\lambda_5,i\lambda_7, \lambda_1,\lambda_3,\lambda_4,
 \lambda_6, \lambda_8)=(T_1,T_2,T_3,T_4,T_5,T_6,T_7,T_8)=(T_I,T_{I'})\ ,
 \ee

 where $\lambda_1,...,\lambda_8$ are the standard Gell-Mann matrices, and
 $I=1,...,p+3, I'=p+4,...,8$ with $0\le p\le 5$. However, the condition
 \eq{c11} is clearly not satisfied since $[T_1,T_2]=T_3$ and thus
 $\hf_{IJ}{}^K\ne 0$. Therefore, we conclude that the
 {\it chiral truncation to a gauged $6D$ theory is not possible
 in this case}.


\subsubsection*{(IV)\ \ $SO(2,1)$}


For this gauging, the minimal $7D$ scalar manifold is
$SO(3,1)/SO(3)$. Let us denote the generators of $SO(3,1)$ by
$T_{AB}=\left(T_{ij},T_{4i}\right)$ where $i=1,2,3$. The $7D$ gauge
group $SO(2,1)$ can be embedded into this $SO(3,1)$ by picking out
the generators $\left(T_{41}, T_{42}, T_{12}\right)$, where the last
generator is compact and the other two are noncompact. Thus,
 \be
 \hf_{\hI\hJ\hK}= \left(g\,\e_{\underline{ijk}}, 0\right)\ ,\quad
 \underline i= 1,2,4\ ,
 \ee

where $\left(T_{41}, T_{42}, T_{12}\right)$ correspond to
$\left(T_1, T_2, T_4\right)$, respectively. The $SO(3,1)$ vector
index, on the other hand, is labeled as $I=1,2,3$ and $I'=4$. Thus,
the conditions \eq{c11} and \eq{ic2} are satisfied and the resulting
$6D$ theory is {\it a $U(1)_R$ gauged supergravity coupled to one
special linear multiplet.} The gauge field is $A_\mu^4$, and the
special linear multiplet lends itself to a description in terms of
the quaternionic Kahler coset $SO(4,1)/SO(4)$. We thus obtain an
{\it $U(1)_R$ gauged supergravity in $6D$ coupled to one
hypermultiplet}. This model is similar to the $Sp(1)_R$ gauged model
obtained from the $SO(3,1)$ gauged $7D$ supergravity described
above, the only difference being that the gauge group is now
$U(1)_R$. In summary, we have the following chain of chiral circle
reduction and hidden symmetry:
 \bea
 {SO(3,1)\over SO(3)}\ \ \lra & (\Phi^i,\varphi)\  \lra  & \ \
 {SO(4,1)\over SO(4)}
\eea

In this case, the  $7D$ theory we start with has $64_B+64_F$ physical
degrees of freedom, while the resulting $6D$ theory has $24_B+24_F$
physical degrees of freedom.


\subsubsection*{(V)\ \ $SO(2,2)$}


This case is of considerable interest as it can be obtained from
a reduction of $N=1$ supergravity in ten dimensions on a certain
manifold $H_{2,2}$ as shown in \cite{Cvetic:2003xr}, where its
chiral circle reduction has been studied. As we shall see
below, their result is a special case of a more general such
reduction.

The minimal model that can accommodate the $SO(2,2)$ gauging is
$SO(3,3)/SO(3)\times SO(3)\sim SL(4,R)/SO(4)$. To solve the conditions
\eq{c11}, we embed the $SO(2,2)$ in $SO(3,3)$ by setting
 \bea
 && {\widehat f}_{\hI\hJ}{}^{\hK}=
 (g_1~\e_{\underline{ij\ell}}~\eta^{\underline{k\ell}}\ ,
 g_2~\e_{\underline{rst}}~\eta^{\underline{tq}}\ )\
 ,\qquad {\underline i}=1,2,6\ ,\quad {\underline r}=3,4,5\ ,
 \nn\w2
 && \eta_{\underline{ij}}={\rm diag}\ (--+)\ ,\qquad
 \eta_{\underline{rs}}={\rm diag}\ (-++)\ ,
 \label{sn}
 \eea

where $(g_1,g_2)$ are the gauge coupling constants for
$SO(2,1)\times SO(2,1)\sim SO(2,2)$. These structure constants can
be checked to satisfy the condition \eq{ic2}. Furthermore, the
conditions \eq{c11} are satisfied since $I=1,2,3,4$ and $I'=5,6$.
The resulting $6D$ theory is {\it a $U(1)_R$ gauged supergravity
coupled to one external Maxwell multiplet (in addition to the
Maxwell multiplet that gauges the R-symmetry) and two
hypermultiplets.} The two hypermultiplets consist of the fields
shown in the last group in \eq{sm} with $p=1, n=3$. The $U(1)_R$ is
gauged by the vector field $A_\mu^6$. The vector field $A_\mu^5$,
which corresponds to $O(1,1)$ rotations, resides in the Maxwell
multiplet. In this model, the surviving $SO(3,1)/SO(3)$ sigma model
sector in $6D$, gets enlarged with the help of the axionic fields to
become the quaternionic Kahler coset $SO(4,2)/SO(4)\times SO(2)$, as
will be shown in section 5. In summary, we have the following chain
of chiral circle reduction and hidden symmetry
 \bea
 {SO(3,3)\over SO(3)\times SO(3)}\ \ \lra\ \ {SO(3,1)\over SO(3)}\ \
 &\lra& \ \ {SO(4,2)\over SO(4)\times
 SO(2)}
 \label{h3}
 \eea

It is also worth noting that the Cvetic-Gibbons--Pope reduction
\cite{Cvetic:2003xr} that gave rise to the $U(1)_R$ gauged 6D
supergravity is a special case of our results that can be obtained
by setting to zero all the scalar fields of the $SO(3,1)/SO(3)$
sigma model, the gauge field $A_\mu^5$ and their fermionic partners.
This model was studied in the language of the $SL(4,R)/SO(4)$ coset
structure. In Appendix B, we give the map between this coset and the
$SO(3,3)/SO(3)\times SO(3)$ coset used here.

Note that, in this case the  $7D$ theory we start with has $64_B+64_F$
physical degrees of freedom, and the resulting $6D$ theory has half
as many, namely, $32_B+32_F$ physical degrees of freedom


\subsubsection*{(VI)\ \ $SO(2,2)\times SO(2,1)$}


In this case, the minimal $7D$ sigma model sector is based on
$SO(6,3)/SO(6)\times SO(3)$. To solve the condition \eq{c11} in such
a way to obtain an $R$-symmetry gauged $6D$ supergravity, we embed
the $7D$ gauge group $SO(2,2)\times SO(2,1)$ in $SO(6,3)$ by setting
 \bea
 && {\widehat f}_{\hI\hJ}{}^{\hK}=
 \left( g_1~\e_{\underline{ij\ell}}~\eta^{\underline{k\ell}}\ ,
 g_2~\e_{\underline{rst}}~\eta^{\underline{tq}}\ ,\ g_3~\e_{i'j'\ell'}
\eta^{k'\ell'} \right)\
 ,\quad
 {\underline i}=1,4,5\ ,\quad {\underline r}=2,6,7\ ,\quad
 i'=3,8,9\ ,
 \nn\w2
 && \eta_{\underline{ij}}={\rm diag}\ (-++)\ ,\qquad
 \eta_{\underline{rs}}={\rm diag}\ (-++)\ ,
 \qquad
 \eta_{i'j'}={\rm diag}\ \ (-++)\ ,
 \eea

where $(g_1,g_2,g_3)$ are the gauge coupling constants for
$SO(2,1)\times SO(2,1)\times SO(2,1)$. The conditions \eq{c11} are
satisfied since $I=1,2,3,5,7,9$ and $I'=4,6,8$. {\it The resulting
$6D$ theory has a local $O(1,1)^3$ gauge symmetry, and hence three
Maxwell multiplets but no gauged $R$ symmetry, and three
hypermultiplets.} The gauge fields are $(A_\mu^4, A_\mu^6,
A_\mu^8)$, and the hypermultiplets consist of the fields shown in
the last group in \eq{sm} with $p=3, n=6$. In this model, the
surviving \ $SO(3,3)/SO(3)\times SO(3)$ sigma model in $6D$ gets
enlarged to the quaternionic Kahler $SO(4,4)/SO(4)\times SO(4)$ with
the help of the axionic fields, as will be described in section 5.
In summary, we have the following chain of chiral circle reduction
and hidden symmetry:
 \bea
 {SO(6,3)\over SO(6)\times SO(3)}\ \
 \lra\ \ {SO(3,3)\over SO(3)\times SO(3)}\ \
 &\lra& \ \ {SO(4,4)\over SO(4)\times SO(4)}
 \label{h4}
 \eea

To summarize, we have found that the $SO(3,1)$ and $SO(2,2)$ gauged
$7D$ models give rise to $U(1)_R$ gauged supergravity, and the
$SO(2,1)$ gauged $7D$ model yields an $Sp(1)_R$ gauged chiral
supergravity, coupled to specific matter multiplets in six
dimensions.


\section{The $6D$ Lagrangian and Supersymmetry Transformations}


 The chiral reduction on a circle along the lines described above requires,
 as usual, the diagonalization of the kinetic terms for various matter fields.
 This is achieved by defining

 \bea
   \sigma &=& \left(  {\hat\sigma}
    -2\alpha\phi\right)\ ,
    \qquad\qquad\qquad\qquad
    \ \ \varphi = \frac12 \left( {\hat\sigma}
     +8\alpha\phi\right)\, ,
     \nn\w2
    \chi &=& {\sqrt 2} \afp\left( {\hat\chi}
     +\frac14{\hat\psi_7}\right)\ ,
    \qquad\quad\quad
    \psi= \frac{1}{\sqrt 2}\,\afp \left(
    {\hat\psi}_7-{\hat\chi}\right)\, ,
     \nn\w2
   \psi_a &=& \frac{1}{\sqrt2}\,\afp \left(
   {\hat\psi}_a-\frac{1}{4}\c_a{\hat\psi}_7\right)\ ,
   \quad\quad\quad\ \ \
    \psi^r=\frac{1}{\sqrt2}\,\afp {\hat\lambda}^r\ ,
    \nn\w2
     \Phi^I &=&{\widehat A}_7^I \ ,\quad\quad
     \lambda^{r'}=\frac{1}{\sqrt2}\,\afp {\hat\lambda}^{r'}\ ,
     \quad\quad\ \ \
     {\widehat\epsilon}= \frac{1}{\sqrt2}\,\afp \epsilon\ .\label{defs8}
 \eea

Furthermore, recalling the Ansatz and noting that the only
non-vanishing components of the spin connection are
 \be
 {\hat\omega}_{cab} =
 e^{-\alpha\phi}\left(\omega_{cab}+2\alpha \eta_{c[a}
 \partial_{b]} \phi\right)\ ,\quad\quad
 \\
 {\hat\omega}_{77a}=\beta e^{-\alpha\phi}  \partial_a \phi\ ,
 \ee

where the indices on the spin connection refer to the tangent space, the
$6D$ supergravity theory obtained by the reduction scheme describe
above has the Lagrangian ${\mathcal L}={\cal L}_B + {\mathcal L}_F$
where\footnote{In order to make contact with more standard
conventions in $6D$, we have redefined $G_{\mu\nu\rho} \ra {\sqrt 2}
G_{\mu\nu\rho}$ and multiplied the Lagrangian by a factor of $1/2$.
The spacetime signature  is $(-+++++)$, the spinors are symplectic
Majorana-Weyl, $C^T=-C$ and $(\c^\mu C)^T=-\c^\mu C$. Thus,
${\bar\psi}\c^{\nu_1\cdots\nu_n}\lambda=
(-1)^n{\bar\psi}\c^{\nu_n\cdots\nu_1}\lambda$, where the $Sp(1)$ doublet indices
are contracted and suppressed. We also use the convention:
$\c_{\mu_1\cdots \mu_6}=e\e_{\mu_1\cdots\mu_6}\,\c_7$.}
 \bea
 e^{-1} \mathcal{L}_B &=&
          \ft{1}{4}R - \frac14 (\p _\mu \sigma)^2
          - \ft{1}{12} e^{2\sigma}
          G_{\mu\nu\rho} G^{\mu\nu\rho}
          - \ft18 e^{\sigma} F_{\mu\nu}^{r'} F^{\mu\nu r'}
   \label{b}\w2
          &&
          -\ft14 \p_\mu\varphi \p^\mu\varphi -\ft14 P_\mu^{ir}P^\mu_{ir}
          -\ft{1}{4} \left(P_\mu^rP^\mu_r +\cP_\mu^i\cP^\mu_i \right)
    \nn\w2
          &&
          -\ft18 e^{-\sigma }\left(
          C^{ir'}C_{ir'} +2S^{ir'}S_{ir'}\right)   \ ,
          \nn\w4
%
%
e^{-1} \mathcal{L}_F &=&
          - \ft{i}{2} {\bar\psi}_\mu
          \c^{\mu\nu\rho}D_\nu\psi_\rho
          - \ft{i}2 {\bar\chi} \c^{\mu} D_\mu \chi
          - \ft{i}{2}{\bar\lambda}^{r'}\c^{\mu} D_\mu\lambda_{r'}
    \label{f}\w2
          &&- \ft{i}{2} {\bar\psi} \c^\mu D_\mu\psi
          - \ft{i}{2}{\bar\psi}^r\c^\mu D_\mu
          \psi^r - \ft{i}2{\bar\chi}\c^\mu\c^\nu\psi_\mu \p_\nu\sigma
    \nn\w2
          && -\ft12 {\bar\psi}^r\c^\mu\c^\nu\sigma_i\psi_\mu P_\nu^{ir}
          +\ft{i}2{\bar\psi}\c^\mu\c^\nu\psi_\mu
          \p_\nu\varphi
    \nn\w2
          && -\ft12{\bar\psi}\c^\mu\c^\nu\sigma_i\psi_\mu \cP_\nu^i
          -{i\over 2}{\bar\psi}^r\c^\mu\c^\nu\psi_\mu P_\nu^r
          -\ft14 \cP_\mu^i X_i^\mu
    \nn\w2
          &&  -iP_\mu^r X_r^\mu
          +\ft{i}{24} e^{\sigma}G_{\mu\nu\rho} X^{\mu\nu\rho}
          - \ft{i}{4} \eps  F^{r'}_{\mu\nu} X^{\mu\nu}_{r'}
    \nn\w2
          && +\es \left( -C_{irr'}{\bar\lambda^{r'}}\sigma^i\psi^r
          +iS_{rr'} {\bar\lambda}^{r'}\psi^r
          - S_{ir'}{\bar\lambda}^{r'}\sigma^i\psi \right)
    \nn\w2
          && +\ft1{2\sqrt 2} \es{\bar\lambda}^{r'}\sigma^i\c^\mu\psi_\mu
          \left( C_{ir'}-{\sqrt 2} S_{ir'}\right)
    \nn\w2
          && + \ft1{2\sqrt 2}\es {\bar\lambda}^{r'}\sigma^i\chi
          \left( C_{ir'}-{\sqrt 2}S_{ir'}\right)\ ,\nn
   \eea

 and where
 \bea
 X^{\mu\nu\rho} &=&
      {\bar\psi}^\lambda \c_{[\lambda}\c^{\mu\nu\rho}
      \c_{\tau]}\psi^\tau
      + {\bar\psi}_\lambda\c^{\mu\nu\rho}
      \c^\lambda \chi
      - {\bar\chi}\c^{\mu\nu\rho}\chi
      + {\bar\lambda}^{r'}\c^{\mu\nu\rho}\lambda_{r'}
      + {\bar\psi}^r\c^{\mu\nu\rho}\psi_r
      +  {\bar\psi}\c^{\mu\nu\rho}\psi\ ,
      \nn\w2
 X^\mu_i &=&
      {\bar\psi}^\rho \c_{[\rho}\c^{\mu}\c_{\tau]}
      \sigma_i\psi^\tau +  {\bar\chi} \c^\mu\sigma_i \chi
      + {\bar\lambda}^{r'} \c^\mu\sigma_i\lambda_{r'}
      - {\bar\psi}^r \c^\mu\sigma_i\psi_r
      -{\bar\psi}\c^\mu\sigma_i\psi\ ,
      \nn\w2
      X^{\mu\nu}_{r'} &=&
      {\bar\psi}_\rho\c^{\mu\nu}\c^\rho\lambda_{r'}
      +  {\bar\chi}\c^{\mu\nu}\lambda_{r'}\ ,
      \nn\w2
 X^\mu_r &=& {\bar\psi}\c^\mu\psi_r\ .
 \label{xdefs}
 \eea

The action is invariant under the following $6D$ supersymmetry
transformations
 \bea
 %
 %
 \delta e_{\mu}^{m} &=&
 i\bar{\epsilon}\c^{m}\psi{}_{\mu}\ ,
 \nn\w2
 \delta \psi_{\mu} &=& D_{\mu} \e -\ft{1}{24}e^{\sigma}
 \c^{\rho\sigma\tau}\c_{\mu}G_{\rho\sigma\tau}\,\e
 -\ft{i}{2} \cP_\mu^i \sigma^i \e\ ,
 \nn\w2
 \delta \chi &=&
 - \ft12 \c^{\mu} \p_{\mu} \sigma\e
 - \ft{1}{12} e^{\sigma}
 \c^{\rho\sigma\tau} G_{\rho\sigma\tau}\,\e \ ,
 \nn\w2
 \delta B_{\mu\nu}&=&
 ie^{-\sigma}\left(\bar{\epsilon}\c_{[\mu}\psi_{\nu]}
 + \ft12 \bar{\epsilon}\c_{\mu\nu} {\chi}\right)
 - A_{[\mu}^{r'} \delta A_{\nu]}^{r'} , \nn\w2
 \delta \sigma &=& -i\bar{\epsilon} \chi\ ,
 \nn\w4
 %
 %
 \delta A_{\mu}^{r'} &=& i e^{- \sigma/2}
 \bar{\epsilon}\c_{\mu} \lambda^{r'}\ ,
 \nn\w2
 \delta \lambda^{r'} &=&
 - \ft{1}{4} \eps \c^{\mu\nu}F_{\mu\nu}^{r'}\e
 -\ft{i}{2\sqrt 2} \es \left( C^{ir'}-{\sqrt 2} S^{ir'}\right)
 \sigma^i\e\ ,
 \nn\w4
 %
 %
 L_I^r\delta \phi^I &=& -i \emf \bar{\epsilon}\psi^r\ ,
 \nn\w2
 %
 %
  L_I^i\delta \phi^I &=& \emf \bar{\epsilon}\sigma^i\psi\ ,
 \nn\w2
 %
 %
 L^I_i\delta L_I^r &=&- \bar{\epsilon} \sigma_i\psi^r\ ,
 \nn\w2
 \delta \varphi &=& i \bar{\epsilon}\psi,
 \nn\w2
 \delta {\psi}&=&
 \ft{i}2 \c^\mu \left(\cP_\mu^i \sigma_i
 -i \p_\mu\varphi \right)\epsilon,
 \nn\w2
 \delta \psi^r &=&
 \ft{i}2 \c^\mu \left( P_\mu^{ir}  \sigma_i +iP_\mu^r\right)\e \ .
 \label{susy2}
 \eea

Several definitions are in order. Firstly, the gauged Maurer-Cartan
form is associated with the coset $SO(p,3)/SO(p)\times SO(3)$ and it
is defined as
 \bea
 P_\mu^{ir} &=& L^{Ii}\left(\p_\mu\delta_I^J - f_{r'I}{}^J A_\mu^{r'} \right)
 L_J^r\ ,\nn\w2
 Q_\mu^{ij} &=& L^{Ii}\left(\p_\mu\delta_I^J - f_{r'I}{}^J A_\mu^{r'}\right)
 L_J^{j}\ ,\nn\w2
 Q_\mu^{rs} &=& L^{Ir}\left(\p_\mu\delta_I^K - f_{r'I}{}^J A_\mu^{r'} \right)
 L_J^{s}\ .
 \label{mc6}
 \eea

Various quantities occurring above are defined as follows:
 \bea
  && G_{\mu\nu\rho}= 3\p_{[\mu} B_{\nu\rho]}
  -\ft32\left( F^{r'}_{[\mu\nu} A_{\rho]}^{r'}
  -\ft13 f_{r's'}{}^{t'}A_\mu^{r'} A_\nu^{s'} A_{\rho t'}\right) \ ,
\nn\w2
  && F_{\mu\nu}^{r'} = 2\p_{[\mu} A_{\nu]}^{r'}
  +f_{s't'}{}^{r'}A_\mu^{s'} A_\nu^{t'}\ ,
\nn\w2
 &&D_\mu \epsilon = \left(\p_\mu
  +\ft14\omega_\mu{}^{ab}\c_{ab}
  +\ft{1}{2\sqrt  2}Q_\mu^i\,\sigma^i\right) \epsilon\ ,
\nn\w2
 && Q_\mu^i=\ft{i}{\sqrt 2}~\epsilon^{ijk}
 Q_{\mu jk}=\epsilon^{ijk}(L^{-1}\p_\mu L)_{jk}+C^{ir'} A_\mu^{r'}\ .
\label{cd6}
 \eea

The axion field strengths are defined as
 \bea
 &&\cP_\mu^i= \ef (D_\mu\phi^I)L_I^i\ , \nn\w2
  &&P_\mu^r=\ef (D_\mu\phi^I)L_I^r\ ,
 \nn\w2
   && D_\mu\Phi^I=\p_\mu\Phi^I+f_{r'J}{}^I\,A_\mu^{r'}\,\Phi^J\ ,
 \label{ax}
 \eea

and the gauge functions as
 \bea
 && C_{kr'}=\ft1{\sqrt 2}\epsilon_{kij} f_{r'I}{}^J\, L^{Ii} L_J^j \ ,
  \quad\quad C_{irr'}= f_{r'I}{}^J L^I_i L_{Jr} \ ,
 \nn\w2
 &&S_{ir'}=-\ef\,f_{r'I}{}^J~\Phi_J\,L^I_i\ ,
 \quad\quad \quad S_{rr'}=-\ef\,f_{r'I}{}^J~\Phi_J\,L^I_r\  .
 \label{gf1}
 \eea

 Note that the $S^2$ term in the potential in \eq{b} comes from the
 $P_7^{ir'}P_7^{ir'}$ term since $P_7^{ir'}\sim S^{ir'}$. The
 above results cover all the chiral reduction schemes that yield
 gauged supergravities in $6D$. We simply need to take the
 appropriate structure constants and the relevant values of $p$ in
 the $SO(p,3)/SO(p)\times SO(3)$ cosets involved.

 In the case of the models (II) and (IV), with $7D$ gauge groups $SO(3,1)$ and
 $SO(2,1)$, respectively, we have $p=0$, which means that the coset
 representative becomes an identity matrix and
 \bea
 && f_{rIJ}\ \ \ra \ \ \e_{r'ij}\ ,\quad\quad\quad\quad
 C_{ijr'}\ \ \ra \ \ {\sqrt 2} \e_{r'ij}\ ,\quad\quad C_{irr'}\ \ \
 \ra \ \ 0\, ,
 \nn\w2
 && S_{ir'}\ \ \ra\ \ -\e_{r'ij} e^{\varphi} \Phi^j\ ,\quad\
 S_{rr'}\ \ \ra\ \ 0\ .
 \label{sp}
 \eea

By an untwisting procedure, which will be described in the next
section, the scalar fields $(\Phi^i,\varphi)$ can be combined to
describe the quaternionic Kahler manifold  $SO(4,1)/SO(4)$ that
governs the couplings of a single hypermultiplet.

In the case of Model (V), we have $p=1$, which means that in the $6D$
model presented above, the relevant sigma model is $SO(3,1)/SO(3)$, while
in the case of  Model (VI), we have $p=2$, which implies the sigma model
$SO(4,2)/SO(4)\times SO(2)$. In each case, the range of indices $I,r,i$
are fixed accordingly.


\section{The Hidden Quaternionic Kahler Coset Structure}


It is well known that the ten dimensional $N=1$ supergravity theory coupled
to $N$ Maxwell multiplets when reduced on a $k$-dimensional torus
down to $D$ dimensions gives rise to half-maximal supergravity
coupled to $(N+k)$ vector multiplets with an underlying
$SO(N+k,k)/SO(N+k)\times SO(k)$ sigma model sector. This means an
$SO(N+3,3)/SO(N+3)\times SO(3)$ sigma model in $7D$. In the notation
of the previous sections, we have $N+3=n$. A circle reduction of
this ungauged theory is then expected to exhibit an
$SO(N+4,4)/SO(N+4)\times SO(4)$ coset structure. This is a well known
phenomenon which has been described in several papers but primarily
in the bosonic sector. In this section, we shall exhibit this
phenomenon in the fermionic sector as well, including  the
supersymmetry transformations. Moreover, we shall describe the
hidden symmetry of the gauged $6D$ models obtained from a consistent
chiral reduction of the gauged $7D$ models, in which case the
$SO(p+3,3)/SO(p+3)\times SO(3)$ coset is enlarged to
$SO(p+4,4)/SO(p+4)\times SO(4)$. Here, we have redefined $p\ra p+3$
compared to the notation of the previous section, for convenience.

The key step in uncovering the hidden symmetry is to first rewrite
the Lagrangian in Iwasawa gauge. This gauge is employed by
parametrizing the coset $SO(p+3,3)/SO(p+3)\times SO(3)\equiv
C(p+3,3)$ by means of the $3(p+3)$ dimensional solvable subalgebra
$K_s$ of $SO(p+3,3)$. The importance of this gauge lies in the fact
that it enables one to absorb the $(p+6)$ axions that come from the
$7D$ Maxwell fields, and a single dilaton that comes from the $7D$
metric, into the representative of the coset $C(p+3,3)$ to form the
representative of the enlarged coset $C(p+4,4)$\footnote{ In
general, the solvable subalgebra ${\widehat K}_s \subset
SO(p+k+1,k+1)$ decomposes into the generators $K_s \subset
SO(p+k,k)$, and $(p+2k)$ generators corresponding to axions and a
single generator corresponding to a dilaton.}. To do so, we shall
first show, in section 5.1, how various quantities {\it formally}
combine to give the enlarged coset structure. This will involve
identifications such as those in \eq{id1} below. These
identifications by themselves do not furnish a proof of the enlarged
coset structure, since one still has to construct explicitly a
parametrization of the enlarged coset which produces these
identifications. In section 5.2, we shall provide the proof by
exploiting the Iwasawa gauge.


\subsection{Hidden Symmetry in the Symmetric Gauge}


The structure of the Lagrangian and transformation rules presented above
readily suggest the identifications $\hP^{ir}=P^{ir}$, $\hQ^{ij}=Q^{ij}$,
$\hQ^{rs}=Q^{rs}$ and
 \bea
  && \pmatrix{\hP^{4r}\cr  \hP^{i,N+4}\cr  \hP^{4,N+4}}
    =\pmatrix{ P^r \cr \cP^i \cr \
    -\p \varphi}\ ,
\quad\quad
  \pmatrix{\hQ^{4i}\cr  \hQ^{N+4,r}}
  =\pmatrix{\cP^i  \cr P^r      }
  \label{id1}
\eea

for the components of the Maurer-Cartan form, $\hC^{ijr'}=C^{ijr'}$,
$\hC^{irr'}=C^{irr'}$ and the following identifications
 \be
 \pmatrix{\hC^{4ir'}\cr \hC^{4rr'}\cr\hC^{i,N+4,r'} \cr \hC^{4,N+4,r'} }
    =\pmatrix{ S^{ir'}\cr S^{rr'}\cr -S^{ir'}\cr 0}\, ,
 \label{cs}
 \ee

where $\hC^{ijr'}=\ft1{\sqrt 2}\,\e^{ijk}~C^{kr'}$, for the gauge
functions. Note that the hat notation here does not refer to higher
dimensions but rather they denote objects which transform under the
enlarged symmetry groups. With these identifications the Lagrangian
simplifies dramatically. The bosonic part takes the form
\bea
 e^{-1} {\mathcal L}_B &=&
          \ft{1}{4}R - \ft14 (\p _\mu \sigma)^2
          - \ft{1}{12} e^{2 \sigma }
          G_{\mu\nu\rho} G^{\mu\nu\rho}
          - \ft18 e^{\sigma} F_{\mu\nu}^{r'} F^{\mu\nu r'}
   \nn\w2
          &&
          - \ft14 \hP_\mu^{\hi\hr} \hP^\mu_{\hi\hr}
           -\ft18 e^{-\sigma} \hC^{\hi\hj r'}\hC_{\hi\hj r'}\ ,
   \label{newb}
   \eea

and the  fermionic part is given by
%
 \bea
   e^{-1} {\mathcal L}_F &=&
   - \ft{i}{2} {\bar\psi}_\mu
          \c^{\mu\nu\rho}D_\nu\psi_\rho
          - \ft{i}2 {\bar\chi} \c^{\mu} D_\mu \chi
          - \ft{i}{2}{\bar\lambda}^{r'}\c^{\mu} D_\mu\lambda_{r'}
          -\ft{i}{2}{\bar\psi}^{\hr}\c^\mu D_\mu\psi^{\hr}
   \w2
          && -\ft{i}2{\bar\chi}\c^\mu\c^\nu\psi_\mu \p_\nu\sigma
             -\ft12 {\bar\psi}^{\hr}\c^\mu\c^\nu{\bar\C}_{\hi}\psi_\mu
             \hP_\nu^{\hi\hr}
             +\ft{i}{24} e^{\sigma}G_{\mu\nu\rho} X^{\mu\nu\rho}
   \nn\w2
          &&
           - \ft{i}{4} e^{\sigma/2} F^{r'}_{\mu\nu} X^{\mu\nu}_{r'}
            - \es \hC_{\hi\hr r'}{\bar\lambda^{r'}}\C^{\hi}\psi^{\hr}
  \nn\w2
         && -\ft{i}4 \es  \hC^{\hi\hj r'}\left({\bar\lambda}^{r'}
         \C_{\hi\hj} \c^\mu\psi_\mu
         +{\bar\lambda}^{r'}\C_{\hi\hj} \chi\right)\ ,
  \label{oxf}
  \eea

where $\hi=1,...,4$, $\hr=1,...,p+4$, we have defined
$\psi^{N+4}=\psi$, and
 \bea
 X^{\mu\nu\rho} &=&
      {\bar\psi}^\lambda \c_{[\lambda}\c^{\mu\nu\rho}
      \c_{\tau]}\psi^\tau
      + {\bar\psi}_\lambda\c^{\mu\nu\rho}
      \c^\lambda \chi
      -  {\bar\chi}\c^{\mu\nu\rho}\chi
      + {\bar\lambda}^{r'}\c^{\mu\nu\rho}\lambda_{r'}
      + {\bar\psi}^{\hr}\c^{\mu\nu\rho}\psi^{\hr}\, ,
 \nn\w2
      X^{\mu\nu}_{r'} &=&
      {\bar\psi}_\rho\c^{\mu\nu}\c^\rho\lambda_{r'}
      +  {\bar\chi}\c^{\mu\nu}\lambda_{r'}\ .
\eea

The covariant derivatives are defined as
\bea
  && D_\mu \pmatrix{\psi_\mu \cr \chi \cr \lambda^{r'}}=
     \left( \nabla_\nu +\ft14\omega_\mu{}^{ab}\c_{ab}
  +\ft14 \hQ_\mu^{\hi\hj}~\C_{\hi\hj}\right)
  \pmatrix{\psi_\nu \cr \chi \cr \lambda^{r'}} \ ,
\nn\w4
 && D_\mu \psi^{\hr}= \left( \p_\mu +\ft14\omega_\mu{}^{ab}\c_{ab}
  +\ft14 \hQ_\mu^{\hi\hj}~{\bar \C}_{\hi\hj} \right) \psi^{\hr}
  +\hQ_\mu^{\hr\hs}~\psi^{\hs}\ .
\eea

The $SO(4)$ Dirac matrices have been introduced in the above formula
with the conventions
 \be
 \C_{\hi}=\left(\sigma_i, -i\right)\ ,\qquad
 {\bar \C}_{\hi}=\left(\sigma_i, i\right)\ ,\qquad
 \C_{\hi\hj}=\C_{[\hi} {\bar \C}_{\hj]}\ ,\qquad
 {\bar \C}_{\hi\hj}={\bar \C}_{[\hi}\C_{\hj]}\ .
 \ee

It is useful to note that ${\bar\psi}^{\hat r}{\bar\C}^{\hi}\e=
-{\bar\e}\C^{\hi}\psi^{\hat r}$.  Looking more closely at
the covariant derivatives
 \bea
  && D_\mu \chi =
     \left( D_\mu(\omega)
  +\ft14 Q_\mu^{ij}~\sigma_{ij} +\ft{i}2 \cP_\mu^i\sigma_i \right) \chi \ ,
  \nn\w4
 && D_\mu \psi^{\hr}= \left( D_\mu(\omega)
  +\ft14 Q_\mu^{ij}~\sigma_{ij} -\ft{i}2 \cP_\mu^i\sigma_i  \right) \psi^{\hr}
  +\hQ_\mu^{\hr\hs}~\psi^{\hs}\ ,
\eea

we observe that they transform covariantly under the composite local
$Sp(1)_R$ transformations inherited from $7D$, and that they contain
the composite $Sp(1)_R$ connections shifted by the {\it positive
torsion} term $(\ft{i}2\cP^i\sigma^i)$ in the case of fermions that
are doublets under the true $Sp(1)_R$ symmetry group in $6D$, namely
$(\psi_\mu, \chi, \lambda^{r'})$, and the {\it negative torsion}
term  $(-\ft{i}2\cP^i\sigma^i)$ in the case of $\psi^{\hr}$, which
are singlets under this symmetry. By true $Sp(1)_R$ symmetry group
in $6D$ we mean the $SO(3)_R$ symmetry group that emerges upon the
recognition of the scalar field couplings as being described by the
quaternionic Kahler coset $SO(p+1,4)/SO(p+1)\times SO(4)$ in which
$SO(4)\sim SO(3)\times SO(3)_R$. The action of this group is best
seen by employing the Iwasawa gauge, as we shall see in the next
subsection.

The action is invariant under the following $6D$ supersymmetry
transformations
 \bea
 %
 %
 \delta e_{\mu}^{m} &=&
 i\bar{\epsilon}\c^{m}\psi{}_{\mu},
 \nn\w2
 \delta \psi_{\mu} &=& D_{\mu} \e -\ft{1}{24}e^{\sigma}
 \c^{\rho\sigma\tau}\c_{\mu}G_{\rho\sigma\tau}\,\e\ ,
 \nn\w2
 \delta \chi &=&
 - \ft12 \c^{\mu} \p_{\mu} \sigma\e
 - \ft{1}{12}e^{\sigma}
 \c^{\rho\sigma\tau} G_{\rho\sigma\tau}\,\e \ ,
 \nn\w2
 \delta B_{\mu\nu}&=&
 ie^{-\sigma}\left(\bar{\epsilon}\c_{[\mu}\psi_{\nu]}
 + \ft12 \bar{\epsilon}\c_{\mu\nu} {\chi}\right)
 - A_{[\mu}^{r'} \delta A_{\nu]}^{r'} , \nn\w2
 \delta \sigma &=& -i\bar{\epsilon} \chi\ ,
 \nn\w4
 %
 %
 \delta A_{\mu}^{r'} &=& i e^{- \sigma/2}
 \bar{\epsilon}\c_{\mu} \lambda^{r'},
 \nn\w2
 \delta \lambda^{r'} &=&
 - \ft{1}{4} \eps \c^{\mu\nu}F_{\mu\nu}^{r'}\e
 +\ft12 \es \hC^{\hi\hj r'}\C_{\hi\hj}\e\ ,
 \nn\w4
 %
 %
 %
 \hL_{\hi}{}^{\hI}\delta \hL_\hI{}^\hr &=& -\bar{\epsilon} \C_{\hi}
 \psi^\hr\ ,
 \nn\w2
 \delta \psi^{\hr} &=&
 \ft{i}2 \c^\mu \hP_\mu^{\hi\hr} {\bar \C}_{\hi}\e \ .
 \eea

 The relation between the supersymmetric variation of the enlarged
 coset representative and those involving the $SO(n,3)/SO(n)\times
 SO(3)$ coset representative, the dilaton and axions is similar to
 the relations in \eq{id1} for the corresponding Maurer-Cartan forms,
 and field strengths, since $L^{-1}dL$ has the same decomposition as
 $L^{-1}\delta L$. Finally, we note that the above results for the
 matter coupled gauged $N=(1,0)$ supergravity in $6D$ are in accordance
 with the results given in \cite{Nishino:1984gk}.


\subsection{Hidden Symmetry in the Iwasawa Gauge}


\subsubsection{The  Ungauged Sector}


In order to comply with the standard notation for the Iwasawa
decomposition of $SO(p,q)$, we switch from our coset representative
to its transpose as $L=\cV^T$. Following
\cite{Lu:1998xt,Cremmer:1997ct}, we then parametrize the coset
$SO(p+3,3)/SO(p+3)\times SO(3)$ as
 \be
 \cV=  e^{\ft12 {\vec \varphi}\cdot {\vec H}}
     e^{C^i{}_j E_i{}^j}~
     e^{\ft12 A_{ij}V^{ij}}~
     e^{B^{i\ur} U_{i\unr}} ~,\ \quad i=1,...,3\ ,\quad \unr =1,...,p\ ,
\ee

where $\left( U_{i\unr}, V^{ij},E_i{}^j, {\vec H}\right)$ with $i<j$
and  $V^{ij}=-V^{ji}$,  are the generators of the $3(p+3)$
dimensional solvable subalgebra of $SO(p+3,3)$ multiplying the
corresponding scalar fields and $\vec\varphi\cdot \vec H$ stands for
$\varphi^i H_i$.

Using the commutation rules of the generators given in Appendix D,
one finds \cite{Lu:1998xt}
 \medskip
 \be {\cV}= \left(
 \begin{array}{c|c|c}
 e^{\fft12\vec c_i\cdot\vec\varphi} \, \c^j{}_i & e^{\fft12\vec
 c_i\cdot\vec\varphi}\, \c^k{}_i\, B^{\uns}{}_k & e^{\fft12\vec
 c_i\cdot\vec\varphi}\,\c^k{}_i\, \left(A_{kj}
 + \ft12 B^{\unq}{}_k\, B^{\unq}{}_j \right) \\
 \hline
 0 & \delta_{\unr\uns} & B^{\unr}{}_j\\
 \hline 0 & 0 & e^{-\fft12\vec c_i\cdot\vec\varphi}\, \wtd\c^i{}_j
 \end{array}\right)\ ,
 \label{cV}
 \ee
 \medskip

where ${\vec c}_i$ is defined in \eq{sc}, and
 \be
 \wtd\c^i{}_j=\d^i_j+C^i{}_j\ ,\qquad \c^i{}_k~\wtd\c^k{}_j~=\d^i_j\ .
 \ee

The inverse of $\cV$ can be computed from the defining relation
$\cV^T\Omega\cV =\Omega$ and is given by:
\medskip
 \be {\cV}^{-1}= \left(
 \begin{array}{c|c|c}
 e^{-\fft12\vec c_j\cdot\vec\varphi} \, \wtd\c^j{}_i &
 -B^{\uns}{}_i & e^{\fft12\vec
 c_j\cdot\vec\varphi}\,\c^k{}_j\, \left(A_{ki}
 + \ft12 B^{\unq}{}_k\, B^{\unq}{}_i \right) \\
 \hline
 0 & \delta_{\unr\uns} & -e^{\fft12\vec c_j\cdot \vec \varphi}\,
 \c^k{}_j\,B^{\unr}{}_k \\
 \hline 0 & 0 & e^{\fft12\vec c_j\cdot\vec\varphi}\,\c^i{}_j
 \end{array}\right)\ .
 \label{cV3}
 \ee
 \medskip

In equations \eq{cV} and \eq{cV3} the indices  $(i,\unr)$ label the
rows and $(j,\uns)$ label the column. The Iwasawa gauge means
setting the scalars corresponding to the maximal compact subalgebra
equal to zero. Under the action of the global $G$ transformations
from the right, the coset representative will not remain in the
Iwasawa gauge but can be brought back to that form by a compensating
$h$ transformation from the left, namely,  $ \cV g=h\cV'$. The
Maurer-Cartan form $d\cV \cV^{-1}$ can be decomposed into two parts
one of which transforms homogeneously under $h$ and the other one
transforms as an $h$-valued gauge field:
 \bea
 P&=& d\cV \cV^{-1}+\left(d\cV \cV^{-1}\right)^T\ :\qquad
 P\ \ \ra \ \ hPh^{-1}\ ,
 \nn\w2
 Q&=&d\cV \cV^{-1}-\left(d\cV \cV^{-1}\right)^T\ :\qquad
 Q\ \ \ra \ \ dh h^{-1} +
 hQh^{-1}\ .
 \label{pq}
\eea

Both of these are, of course, manifestly invariant under the global
$g$ transformations. The key building block in writing down the
action in Iwasawa gauge is the Maurer-Cartan form \cite{Lu:1998xt}
 \be
 \partial_\mu \cV \cV^{-1}=
 \ft12 \p_\mu{\vec\varphi}\cdot {\vec H}
 +\sum_{i<j}\left(  e^{\fft12{\vec a}_{ij}\cdot{\vec\varphi}}
  F_{\mu ij} V^{ij}+  e^{\fft12{\vec b}_{ij}\cdot{\vec\varphi}}
  \cF_\mu^i{}_jE_i{}^j\right)+\sum_{i,\unr} F_{\mu i\unr} U^{i\unr}\
  ,
\label{mc4}
 \ee

where ${\vec a}_{ij}$ and ${\vec b}_{ij}$ are defined in \eq{sc} and
 \bea
 F_{\mu ij}&=&\c^k{}_i\c^\ell{}_j\left(\p_\mu A_{k\ell}
 -B^{\unq}{}_k\p_\mu B^{\unq}{}_\ell\right)\ ,
  \nn\w2
  F_{\mu i\unr}&=& \c^j{}_i\p_\mu B_{j\unr}\ ,
  \nn\w2
  \cF_\mu^i{}_j &=& \c^k{}_j \p_\mu C^i{}_k\ ,
  \label{did}
  \eea

and it is understood  that $i<j$. Other building blocks for the
action are the field strengths for the axions defined as
 \be
 \cV  \p_\mu \Phi =\pmatrix{F_{\mu i} \cr F_\mu^{\unr} \cr \cF_\mu^i}\ ,
 \quad\quad \Phi =\pmatrix{A_i \cr B^{\unr} \cr C^i}\ ,
\label{ax2}
 \ee

where $A_i$ and $B^{\ur}$ form a $(p+3)$ dimensional representation
of $SO(p+3)\subset SO(p+3,3)$, and
 \bea
 F_{\mu i}&=& \c^j{}_i\left(\p_\mu A_j+ B^{\unr}{}_j\p_\mu B^{\unr}
 +A_{jk}\p_\mu C^k +\ft12 B^{\unr}{}_j B^{\unr}{}_k\p_\mu C^k
\right)\ ,
 \nn\w2
 F_\mu^{\unr} &=& \p_\mu B^{\unr} +B^{\unr}{}_i\p_\mu C^i\ ,
 \nn\w2
 \cF_\mu^i&=& \wtd\c^i{}_j~\p_\mu C^j\ .
 \label{axd}
 \eea

With these definitions, the bosonic part of our $6D$ Lagrangian that
contains the $(4p+4)$ scalar fields, which we shall call
$\cL_{G/H}$, takes the form
 \bea
 e^{-1}{\cL}_{G/H} &=&
 -\ft14 \p_\mu\varphi\p^\mu\varphi -\ft18 {\rm tr}~
 \left(d\cV\cV^{-1}\right)\left( d\cV\cV^{-1}+ ( d\cV\cV^{-1})^T\right)
 \nn\w2
 && -\ft14 e^{2\varphi} (\cV \p_\mu \Phi)^T (\cV \p^\mu \Phi) \ .
\eea

More explicitly, this can be written as \cite{Lu:1998xt}

 \bea
 e^{-1}{\cL}_{G/H}&=&
 -\ft14 \p_\mu\varphi\p^\mu\varphi -\ft18 \sum_{i<j}
 \left(e^{{\vec a}_{ij}\cdot{\vec\varphi}}F_{\mu ij}F^{\mu ij}
 +e^{{\vec b}_{ij}\cdot{\vec\varphi}}{\cF_\mu}^i{}_j {\cF^\mu}_i{}^j \right)
 -\sum_{i,r}\ft18 e^{{\vec c}_i\cdot{\vec\varphi}}   F_{\mu i\unr}F^{\mu i\unr}
 \nn\w2
 &&-\ft14 e^{2\varphi}\left( e^{{\vec c}_i\cdot{\vec\varphi}}F_{\mu i}
 F^{\mu i}+ e^{-{\vec c}_i\cdot{\vec\varphi}}
 \cF_\mu^i\cF^\mu_i+F_\mu^{\unr} F^\mu_{\unr}\right)\ .
 \label{gh1}
 \eea

The idea is now to combine the dilaton and axionic scalar field
strengths \eq{axd} with the scalar field strengths for
$SO(p+3)/SO(p)\times SO(3)$ defined in \eq{did} to express them all
as the scalar field strengths of the enlarged coset
$SO(p+4,4)/SO(p+4)\times SO(4)$. As is well known, this is indeed
possible and to this end we need to make the identifications
 \bea
 F_{\mu i} &=& \ft1{\sqrt 2}\,F_{\mu i4}\ ,
 \nn\w2
 F_\mu^{\unr} &=& \ft1{\sqrt 2}\,F_{\mu 4}^{\unr}\ ,
 \nn\w2
 \cF_{\mu}^i &=& \ft1{\sqrt 2}\,\cF_\mu^i{}_4\ .
 \label{id3}
 \eea

The quantities on the right hand side are restrictions of the
Maurer-Cartan form based on the enlarged coset
$SO(p+4,4)/SO(p+4)\times SO(3)$ defined as

 \be
 \p_\mu \hcV \hcV^{-1} =
 \ft12 \p_\mu \varphi^\a H_\a
 +\sum_{\a<\b}\left(  e^{\fft12{\vec a}_{\a\b}\cdot{\vec\varphi}}
  F_{\mu \a\b} V^{\a\b}+  e^{\fft12{\vec b}_{\a\b}\cdot{\vec\varphi}}
  \cF_\mu^\a{}_\b E_\a{}^\b\right)+\sum_{\a,r}
  e^{\fft12{\vec c}_\a\cdot{\vec\varphi}}F_{\mu \a \unr} U^{\a \unr}\ ,
\label{mc5}
  \ee

where  ${\hcV}$ is defined as in \eq{cV} and $(F_{\mu \a\b},
 \cF_\mu^\a{}_\b, F_{\mu \a \unr})$ as in \eq{did}, and ${\vec a}_{\a\b},
 {\vec b}_{\a\b}, {\vec c}_\a$ as in \eq{sc}, with the
 {\it 3-valued indices replaced by the 4-valued indices everywhere}.
Equations \eq{id3} have a solution given by
 \bea
 A_i &=& \ft1{\sqrt 2}\left(
 A_{i4}-A_{ij} \c^j{}_k C^k{}_4 -\ft12B^{\unr}{}_iB^{\unr}{}_4
 +\ft12 B^{\unr}{}_i B^{\unr}{}_j\c^j{}_k C^k{}_4\right)\ ,
 \nn\w2
 B^{\unr} &=& \ft1{\sqrt 2}\left( B^{\unr}{}_4
 -B^{\unr}{}_i \c^i{}_jC^j{}_4\right)\
 ,
 \nn\w2
 C^i &=& \ft1{\sqrt 2}\c^i{}_j C^j{}_4\ ,
 \nn\w2
 \varphi &=& \ft1{\sqrt2}\,\varphi_4\ .
 \label{sol5}
 \eea

The identifications \eq{id3} (where $r\ra \ur, i$), together with
\eq{ax}, \eq{ax2}, \eq{mc5}, \eq{mc4}, \eq{mc6} (with $A_\mu=0$),
\eq{pq} and \eq{sc} (defined for 3-valued and 4-valued indices
similarly), provide the proof of \eq{id1} used to show the hidden
symmetry. Using \eq{id3}, the Lagrangian ${\mathcal L}_{G/H}$ can be
written as the $SO(p+4,4)/SO(p+4)\times SO(4)$ sigma model:
 \bea
 e^{-1}{\cL}_{G/H}&=&
 -\ft18  {\rm tr}~
 \left(d{\widehat\cV}{\widehat\cV}^{-1}\right)
 \left( d{\widehat\cV}{\widehat\cV}^{-1}
 + ( d{\widehat\cV}{\widehat\cV}^{-1})^T\right)
 \w3
 &=& -\ft18 \p\varphi^\a\p^\mu \varphi_\a
 -\ft18\sum_{\a<\b}
 \left(e^{{\vec a}_{\a\b}\cdot{\vec\varphi}}F_\mu^{\a\b}F^\mu_{\a\b}
 +e^{{\vec b}_{\a\b}\cdot{\vec\varphi}}
 {\cF_\mu}_\a{}^\b {\cF^\mu}^\a{}_\b\right)
 -\ft18\sum_{\a,r} e^{{\vec c}_\a\cdot{\vec\varphi}}
 F_{\mu\a \unr}F^{\mu\a \unr}\ ,\nn
 \eea

where $a_{\a\b}, b_{\a\b}, c_\a$ are defined as in \eq{sc} with the
indices $i,j=1,2,3$ replaced by $\a,\b=1,...,4$.


\subsubsection{Gauging and the C-functions}


In order to justify the identifications \eq{cs} of the $S$-functions
as certain components of the $C$-functions associated with the
enlarged coset space, we need to study these functions in the
Iwasawa gauge. Of the four supergravity models in $6D$ that have
nonvanishing gauge functions, two of them, namely models (II) and
(IV), see section 3.2, are coupled to one special linear multiplet
and as such they deserve separate treatment. In both of these cases,
the gauge functions obey the relations \eq{sp}. We begin by showing
how these relations follow from the $C$-functions associated with
the $U(1)_R$ or $Sp(1)_R$ gauged $SO(4,1)/SO(4)$ sigma model.

The coset representative for $SO(4,1)/SO(4)$ in the Iwasawa gauge
takes the form
 \medskip
 \be
 {\cV}= \left(
 \begin{array}{c|c|c}
 e^\varphi  & e^\varphi \Phi^i & \ft12 e^\varphi \Phi^2   \\
 \hline
 0 & \delta_{ij} & \Phi^i\\
 \hline 0 & 0 & e^{-\varphi}
 \end{array}\right)\ ,
 \label{cV2}
 \ee

where $\Phi^2=\Phi^i \Phi_i$. Note that we have made the
identification $B_{1r}\ra \Phi_i$ already. Given that the $Sp(1)_R$
or $U(1)_R$ generators $T^{r'}$ are of the form
 \medskip
 \be
 T^{r'}= \left(
 \begin{array}{c|c|c}
 0 & 0 & 0 \\
 \hline
 0 & T^{r'} & 0\\
 \hline 0 & 0 & 0
 \end{array}\right)\ ,
 \label{gc2}
 \ee

we find that the $C$ function based on the enlarged coset is given by
 \be
 C^{r'}= {\cV}T^{r'}{\cV}^{-1}=\left(
 \begin{array}{c|c|c}
 0 & - T^{r'}_{ij}\Phi^j & 0 \\
 \hline
 0 & T^{r'}_{ij} & - T^{r'}_{ij}\Phi^j\\
 \hline 0 & 0 & 0
 \end{array}\right)\ .
 \label{gc3}
 \ee

Comparing with the relations given in \eq{sp}, and recalling that
$T^{r'}_{ij} \sim \e_{r'ij}$, we see that indeed the projection of
the $C$ function based on the enlarged coset as defined above does
produce the $C$ and $S$ functions obtained from the chiral
reduction, as was assumed in the previous section in \eq{cs}. In the
notation of Appendix D, the $C$ function is obtained from projection
by $T_{ij}$, and the $S$-function from projection by $U_i$.

Next, we consider the remaining two supergravities with nontrivial
gauge function, namely models (V) and (VI), see section 3.2, with
hidden symmetry chains shown in \eq{h3} and \eq{h4}. Prior to
uncovering the hidden symmetry, the C and S functions occurring in
the Lagrangian \eq{b} and in the supersymmetry transformations
\eq{susy2} are $C^{ijr'}, C^{irr'}, S^{ir'}$ and $S^{rr'}$. Using
\eq{gf1}, and the fact that $f_{r'I}{}^J \sim (T_{r'})_I{}^J$, we
deduce the definitions
 \bea
 {\vec C}^{r'} (H)&=& \tr\,
 \left( {\cV}T^{r'}{\cV}^{-1}\right)\,{\vec H}\ ,
 \nn\w2
 C_i{}^{jr'} (E)&=& \tr\, \left(
 {\cV}T^{r'}{\cV}^{-1}\right)\,E_i{}^j\ ,
 \nn\w2
 C_{ij}{}^{r'} (V)&=& \tr\, \left(
 {\cV}T^{r'}{\cV}^{-1}\right)\,V^{ij}\ ,
 \nn\w2
 C_{i\unr}{}^{r'}(U) &=& \tr\, \left( {\cV} T^{r'}{\cV}^{-1}\right)
 \,U_{i\unr}\ ,
 \label{cfn}
 \eea

for the $C$ functions in the coset direction,
 \be
 C^{ijr'} (X)= \tr\,\left({\cV} T^{r'}{\cV}^{-1}\right)\, X^{ij}\ ,
 \label{cfn1}
 \ee

for the $C$ function in the $SO(3)$ direction, and
 \be
 S^{r'}=\pmatrix{{\cal S}^{r'}_i \cr S^{rr'} \cr S^{ir'}}
 ={\sqrt 2}\, e^{\varphi}\,{\cV}T^{r'}\Phi\ ,
\label{cfn2}
 \ee

for the $S$ functions. For $SO(p,q)/SO(p)\times SO(q)$ with $p\ge
q$, we have $i=1,...,q$ and $\unr =1,...,p-q$. To show that the $C$
and $S$ functions defined above combine to give the ${\hat C}$
functions for the enlarged coset, we begin with the observation that
the gauge group lies in the $H^i$ and $T_{\unr\uns}$ directions. Thus we
can denote the full gauge symmetry generator that acts on the
enlarged coset representative $\cV $ as
 \medskip
 \be
 T^{r'}= \left(
 \begin{array}{c|c|c}
  H^{r'}& 0 & 0 \\
 \hline
 0 & T^{r'} & 0\\
 \hline 0 & 0 & H^{r'}
 \end{array}\right)\ ,
 \label{gc22}
 \ee

where $H^{r'}_{\a\b}\ (\a,\b=1,...,q+1)$ is symmetric and
$T^{r'}_{rs} (r,s =1,...,p-q)$ is antisymmetric. Defining the $C$
functions for the coset $SO(p+1,q+1)/SO(p+1)\times SO(q+1)$ as in
\eq{cfn} with the index $i=1,...,q$ replaced by $\a=1,...,q+1$ and
$T^{r'}$ defined in \eq{gc22}, we find
 \bea
 {\vec C}^{r'}(H)
 &=& \c^\c{}_\a\,H^{r'}_\c{}^\d\,{\tilde\c}^\a{}_\d\, {\vec c}_\a\ ,
 \nn\w2
 C^\b{}_\a{}^{r'}(E)
 &=&e^{\fft12({\vec c}_\a-{\vec c}_\b)\cdot {\vec
 \varphi}}\,\c^\c{}_\a\,H^{r'}_\c{}^\d\,{\tilde\c}^\b{}_\d\ ,
 \nn\w2
{\cal C}_{\a\b r'}(V)&=&
 e^{\fft12({\vec c}_\a-{\vec c}_\b)\cdot {\vec\varphi}}\,
 \c^\c{}_{[\a}\, \c^\d{}_{\b]}\left(H^{r'}_\c{}^\eta
 ( A_{\d\eta}+\ft12 B_\d^{\unr} B_\eta^{\unr})
 -T^{r'}_{\unr\uns} B_\c^{\unr} B_\d^{\uns}\right)\ ,
 \nn\w2
 C_{\a \unr}{}^{r'} (U)&=& -e^{\fft12{\vec c}_\a
 \cdot {\vec\varphi}}\,\c^\b{}_\a
 \left(H^{r'}_\b{}^\c\,B_{\c \unr}+T^{r'}_{\unr\uns}
 B_\b{}^{\uns}\right)\ .
 \label{ch3}
 \eea

Using \eq{sol5}, the above quantities reduce to those for the
$SO(p+3,3)/SO(p+3)\times SO(3)$ coset upon restriction of the
 4-valued ${\a,\b}$ indices to 3-valued $(i,j)$ indices,
 $ C^{4r'}(H)=0$ and
 \bea
 C^{r'}_{4i}(E)&=& {\sqrt 2} e^{\varphi} {\cal S}^{r'}{}_i\ ,
 \nn\w2
 C^{r'}_{4i}(V)&=& {\sqrt 2} e^{\varphi} S^{r'}{}_i\ ,
 \nn\w2
 C^{r'}_{4r}(U)&=& -{\sqrt 2} e^{\varphi} S^{r'}{}_r\ .
 \eea

These identifications, upon comparing the definitions \eq{cfn},
\eq{cfn1} and \eq{cfn2} with \eq{gf1}, provide the proof of the
relations \eq{cs} used in showing the hidden symmetry. In doing so,
note that $C^{irr'}\ra (C^{i\unr r'}, C^{i,j r'})$ and $S^{rr'} \ra
(S^{\unr r'}, {\cal S}^{ir'})$ and that $C^{i,j r'}$ has components
in the $({\vec H}, E, V, U,X)$ directions. Note also that, having proven
the relations \eq{id1} and \eq{cs}, it follows that not only the
bosonic part of the $6D$ Lagrangian, \eq{newb}, but also
its part that contains the fermions, namely,
\eq{oxf}, exhibits correctly the enlarged coset structure.
This concludes the demonstration of the enlarged coset structure in the
$6D$ models with nontrivial gauge functions.


\section{Comments}


We have reduced  the half-maximal $7D$ supergravity with specific
noncompact gaugings coupled to a suitable number of vector
multiplets on a circle to $6D$ and chirally truncated it to
$N=(1,0)$ supergravity such that a $R$-symmetry gauging survives.
These are referred to as the $SO(3,1), SO(3,1)$ and $SO(2,2)$
models, and their field content and gauge symmetries are summarized
in the Introduction. These models, in particular, feature couplings
to $p$ linear multiplets whose scalar fields parametrize the coset
$SO(p,3)/SO(p)\times SO(3)$, a dilaton and $(p+3)$ axions, for $p\le
1$. The value of $p$ is restricted in the case of chiral circle
reductions that maintain $R$-symmetry gauging, but it is arbitrary
otherwise. We have exhibited in the full model, including the
fermionic contributions, how these fields can be combined to
parametrize an enlarged coset $SO(p+1,4)/SO(p+1)\times SO(4)$ whose
abelian isometries correspond to the $(p+3)$ axions. In the ungauged
$6D$ models obtained, we have dualized the axions to $4$-form
potentials, thereby obtaining the coupling of $p$ linear multiplets
and one special linear multiplet to chiral $6D$ supergravity (see
Appendix C).


In this paper, we have also shown that, contrary to the claims made
in the literature \cite{Park:1988id}, the 2-form potential in the
gauged $7D$ supergravity can be dualized to a 3-form potential even
in the presence of couplings to an arbitrary number of vector
multiplets.

Our results for the $R$-symmetry gauged reduction of certain
noncompact gauged $7D$ supergravities are likely to play an
important role in finding the string/M-theory origin of the {\it
gauged and anomaly-free} $N=(1,0)$ supergravities in $6D$ which has
been a notoriously challenging problem so far. This is due to the
fact that at least two of the $7D$ models we have encountered, namely the
$SO(3,1)$ and $SO(2,2)$ gauged $7D$ models, are known to have a
string/M-theory origin. Therefore, what remains to be understood is
the introduction of the matter couplings in $6D$ that are needed for
anomaly freedom. A natural approach for achieving this to associate
our chiral reduction with boundary conditions to be imposed on the
fields of the $7D$ model formulated on a manifold with boundary
\cite{Avramis:2004cn}.


\section*{Acknowledgments}


We are grateful to Mees de Roo, P. Howe, Ulf Lindstr\"om, Hong Lu,
Chris Pope, Seif Randjbar-Daemi, Kelly Stelle and Stefan Vandoren
for discussions. E.S. thanks the Institute for Theoretical Physics
at Groningen, and at Uppsala, where part of this work was done, for
hospitality. The work of E.S. and D.J. Jong is supported in part by
NSF grant PHY-0314712. The work of E.B. was supported by the EU
MRTN-CT-2004-005104 grant `Forces Universe' in which E.B. is
associated to Utrecht University.

\newpage

\begin{appendix}


\section{The Dual Gauged 7D Model with Matter Couplings and
Topological Mass Term}


The 2-form potential occurring in the model of \cite{Bergshoeff:1985mr}
given above can easily be dualized \cite{Avramis:2004cn} to a 3-form
potential. To do this, one adds the total derivative term to obtain the
new Lagrangian
 \be
 {\cal L}_3= {\mathcal L} -\ft1{144} \epsilon^{\mu_1\cdots \m_7} H_{\mu_1\cdots
 \mu_4} \left(G_{\mu_5\cdots\mu_7}
 +\ft3{\sqrt 2}\omega^0_{\mu_5\cdots \mu_7}\right)\ ,
 \ee

 where
 \be H_{\mu\nu\rho\sigma}= 4\p_{[\mu}C_{\nu\rho\sigma]}\ .
 \ee

We can treat $G$ as an independent field because the $C$-field equation
will impose the correct Bianchi identity that implies the correct form of
$G$ given in the previous section. Thus, treating $G$ as an independent
field, its field equation gives
 \be
 G_{\mu\nu\rho}=-\ft1{24}e^{-2\sigma}
 e\epsilon_{\mu\nu\rho\sigma_1\cdots\sigma_4} H^{\sigma_1\cdots\sigma_4}
 +\ft{i}{4\sqrt 2} e^{-\sigma}X_{\mu\nu\rho}\ .
 \label{3g}
 \ee

Using this result in the Lagrangian given in \eq{2t}, one finds
 \bea
 {\cal L}_3 &=& {\cal L}'
 -\ft1{48}ee^{-2\sigma} H_{\mu\nu\rho\sigma}H^{\mu\nu\rho\sigma}
 -\ft1{48\sqrt 2}\epsilon^{\mu_1\cdots \mu_7}
 H_{\mu_1\cdots \mu_4} \omega^0_{\mu_5\cdots\mu_7}
 \nn\w2
 &&-\ft{i}{576\sqrt 2}
 e^{-\sigma} \epsilon^{\mu_1\cdots\mu_7}
 H_{\mu_1\cdots \mu_4} X_{\mu_5\cdots \mu_7}\ ,
 \label{dual}
 \eea

where ${\cal L}'$ is the $G$-independent part of \eq{2t}. For the
readers convenience, we explicitly give the dual Lagrangian
${\cL}_3=\cL_{3B}+{\cL}_{3F}$ where
 \bea
 e^{-1}{\cal L}_{3B} &=& \ft12 R -\ft14 e^\sigma a_{IJ}F_{\mu\nu}^I
 F^{\mu\nu J}
 -\ft1{48}ee^{-2\sigma} H_{\mu\nu\rho\sigma}H^{\mu\nu\rho\sigma}
 -\ft1{48\sqrt 2}\epsilon^{\mu_1\cdots \mu_7}
 H_{\mu_1\cdots \mu_4} \omega^0_{\mu_5\cdots\mu_7}\nn\\
 && -\ft58
 \p_\mu\sigma \p^\mu \sigma - \ft12 P_\mu^{ir}
 P^\mu_{ir}-\ft14 e^{-\sigma} \left(C^{ir} C_{ir}
 -\ft19 C^2\right)\ ,
 \w4
 e^{-1}{\cal L}_{3F} &=&-\ft{i}2
 {\bar\psi}_\mu\c^{\mu\nu\rho}D_\nu\psi_\rho -\ft{5i}2
 {\bar\chi}\c^\mu D_\mu\chi -\ft{i}2 {\bar\lambda}^r
 \c^\mu D_\mu\lambda_r -\ft{5i}4 {\bar\chi}
 \c^\mu\c^\nu\psi_\mu \p_\nu\sigma -\ft12
 {\bar\lambda}^r \sigma^i\c^\mu\c^\nu\psi_\mu P_{\nu ri}
 \nn\w2
 &&+\ft{i}{96\sqrt 2} e^\sigma H_{\mu\nu\rho\sigma} X^{\mu\nu\rho\sigma}
 +\ft18 e^{\sigma/2} F_{\mu\nu}^i X_i^{\mu\nu}
 -\ft{i}{4} e^{\sigma/2} F_{\mu\nu}^r X_r^{\mu\nu}
 \nn\w2
 &&-\ft{i\sqrt 2}{24} e^{-\sigma/2} C\left(
 {\bar\psi}_\mu\c^{\mu\nu}\psi_\nu
 +2{\bar\psi}_\mu\c^\mu\chi+3{\bar\chi}\chi-{\bar\lambda}^r\lambda_r
 \right)
 \nn\w2
 &&+\ft1{2\sqrt 2} e^{-\sigma/2} C_{ir}
 \left( {\bar\psi}_\mu \sigma^i\c^\mu\lambda^r
 -2{\bar\chi}\sigma^i\lambda^r\right)+\ft12
 e^{-\sigma/2}C_{rsi} {\bar\lambda}^r\sigma^i\lambda^s\ ,
 \eea

and where the fermionic bilinears are defined as
 \bea
 X^{\mu\nu\rho\sigma}&=& {\bar\psi}^\lambda\c_{[\lambda}
 \c^{\mu\nu\rho\sigma}\c_{\tau]} \psi^\tau
 +
 4{\bar\psi}_\lambda\c^{\mu\nu\rho\sigma}\c^\lambda\chi
 -3{\bar\chi}\c^{\mu\nu\rho\sigma}\chi
 +{\bar\lambda}^a\c^{\mu\nu\rho\sigma}\lambda_a\ ,
 \nn\w2
 X^{i\mu\nu}&=& {\bar\psi}^\lambda \sigma^i \c_{[\lambda}
 \c^{\mu\nu}\c_{\tau]} \psi^\tau
 -2{\bar\psi}_\lambda\sigma^i\c^{\mu\nu}\c^\lambda\chi
 +3{\bar\chi}\sigma^i\c^{\mu\nu}\chi
 -{\bar\lambda}^r\sigma^i\c^{\mu\nu}\lambda_r\ ,
 \nn\w2
 X^{r\mu\nu}&=& {\bar\psi}_\lambda
 \c^{\mu\nu}\c^\lambda\lambda^r
 +2{\bar \chi}\c^{\mu\nu}\lambda^r\ .
 \eea

 The supersymmetry transformation rules are
\bea
 \delta e_\mu{}^m &=& i\eb \c^m\psi_\mu\ ,
 \nn\w2
 \delta \psi_\mu &=& 2D_\mu\epsilon
 -\ft{\sqrt 2}{30} e^{-\sigma/2}C\c_\mu\epsilon
 \nn\\
 &&-\ft1{240\sqrt 2}e^{-\sigma}
 H_{\rho\sigma\lambda\tau}\left(\c_\mu\c^{\rho\sigma\lambda\tau}
 +5\c^{\rho\sigma\lambda\tau}\c_\mu \right)\epsilon
 -\ft{i}{20} e^{\sigma/2} F_{\rho\sigma}^i\, \sigma^i
 \left(3\c_\mu\c^{\rho\sigma}-5\c^{\rho\sigma}\c_\mu\right)
 \epsilon\, ,
 \nn\w3
 \delta \chi&=&-\ft12\c^\mu\p_\mu\sigma\epsilon
 -\ft{i}{10} e^{\sigma/2}F_{\mu\nu}^i\,\sigma^i \c^{\mu\nu}\epsilon
 -\ft1{60\sqrt 2}
 e^{-\sigma} H_{\mu\nu\rho\sigma}\c^{\mu\nu\rho\sigma}\epsilon
 +\ft{\sqrt 2}{30}e^{\sigma/2}C\epsilon\ ,
 \nn\w2
 \delta C_{\mu\nu\rho} &=& e^\sigma\left(  \ft{3i}{\sqrt 2}\eb
 \c_{[\mu\nu}\psi_{\rho]} -i{\sqrt 2}
 \eb\c_{\mu\nu\rho}\chi\right) \ ,
 \label{3s}\w2
 \delta \sigma &=& -2i\eb\chi\ ,
 \nn\w2
 \delta A_\mu^I &=& -e^{-\sigma/2}\left(\eb\sigma^i\psi_\mu
 +\eb\sigma^i\c_\mu\chi\right) L^I_i
 +ie^{-\sigma/2} \eb\c_\mu\lambda^r L^I_r\ ,
 \nn\w2
 \delta L_I^r &=& \eb\sigma^i \lambda^r L_I^i\ ,\quad\quad
 \delta L_I^i = \eb\sigma^i \lambda_r L_I^r\ ,
 \w2
 \delta \lambda^r &=& -\ft12 e^{\sigma/2}F_{\mu\nu}^r
 \c^{\mu\nu}\epsilon +i\c^\mu P_\mu^{ir}\sigma^i\epsilon
 -\ft{i}{\sqrt 2} e^{-\sigma/2} C^{ir} \sigma^i \epsilon\ .
 \nn
 \eea

The supersymmetry transformation rule for the 3-form potential can
be obtained from the supersymmetry of the $G$-field equation
\eq{3g}. Indeed, it is sufficient to check the cancellation of the
$\p_\mu \epsilon $ terms to determine the supersymmetry variation of
the 3-form potential. If we set to zero all the vector multiplet
fields, the above Lagrangian and transformation rules become those of
$SU(2)$ gauged pure half-maximal supergravity
\cite{Townsend:1983kk}, which in turn admits a topological mass term
for the 3-form potential in a supersymmetric fashion that involves a
new constant parameter \cite{Townsend:1983kk}. In \cite{Park:1988id},
it has been argued that the gauged theory in presence of the
coupling to vector
multiplets does not admit a topological mass term. However, we have
found that this is not the case. Indeed, we have found that one can
add the following Lagrangian to ${\cal L}_3$ given in \eq{dual}:
 \bea
 e^{-1}{\cal L}_h &=& {he^{-1}\over 36}\,\epsilon^{\m_1\cdots \mu_7}
 H_{\mu_1\cdots \mu_4} C_{\mu_5\cdots \mu_7}
 +\ft{4\sqrt 2}{3} he^{3\sigma/2} C -16ih^2 e^{4\sigma}\nn\w2
 &&ihe^{2\sigma} \left(-{\bar\psi}_\mu\c^{\mu\nu}\psi_\nu
 +8{\bar\psi}_\mu\c^\mu \chi +27{\bar\chi}\chi-{\bar\lambda}^r
 \lambda_r\right)\ .
 \label{Lh}
 \eea

Note that the coupling of matter to the model with topological mass term
has led to the dressing up of the term $he^{3\s/2}$ present in that
model by $C$ as shown in the second term  on the right hand side of
\eq{Lh}.  The second ingredient to make the supersymmetry work is
the term $h e^{2\s}{\bar\lambda}^r\lambda_r$ in \eq{Lh}\footnote{The
obstacle reported in \cite{Park:1988id} in coupling matter in
presence of the topological terms may be due to the fact that these
ingredients were not considered.}. The action for the total
Lagrangian
 \be
 {\cal L}_{new}= {\cal L}_3 + {\cal L}_h
 \label{3new}
 \ee

is invariant under the supersymmetry transformation rules described above
with the following new $h$-dependent terms:
 \bea
 \delta_h \psi_\mu &=& -\ft45 he^{2\sigma}\c_\mu\epsilon\ ,
 \nn\w2
 \delta_h \chi &=& -\ft{16}5 he^{2\sigma}\epsilon\ .
 \label{3h}
 \eea

For comparison with \cite{Townsend:1983kk}, we extract the potential
and all the mass terms, and write it as
 \bea
 \Delta{\cal L}&=& 60 m^2-10\left( m+2he^{2\sigma}\right)^2
 +\ft{5im}{2} {\bar\psi}_\mu\c^{\mu\nu}\psi_\nu
 -5i\left( m+2he^{2\sigma}\right){\bar\psi}\c^\mu\chi
 \nn\w2
 &&+5i\left(\ft32 m+6he^{2\sigma}\right) {\bar\chi}\chi
 -\ft{i}2 \left(5m+4he^{2\sigma}\right){\bar\lambda}^r\lambda_r
 -\ft14 e^{-\sigma} C^{ir} C_{ir}\nn\w2
 &&+\ft1{2\sqrt 2} e^{-\sigma/2} C_{ir}
 \left( {\bar\psi}_\mu \sigma^i\c^\mu\lambda^r
 -2{\bar\chi}\sigma^i\lambda^r\right)+\ft12
 e^{-\sigma/2}C_{rsi} {\bar\lambda}^r\sigma^i\lambda^s\ ,
 \eea

where we have defined
 \be
 m=-\ft{1}{30\sqrt 2}\,C e^{-\sigma/2} -\ft25 h e^{2\sigma}\ ,
 \ee

 so that
 \bea
 \delta' \psi_\mu &=& 2m\c_\mu \epsilon\ ,
 \nn\w2
 \delta'\chi &=& -2(m+2he^{2\sigma}) \epsilon\ .
 \eea

In the absence of matter couplings, the above result has exactly the
same structure as that of \cite{Townsend:1983kk} but the
coefficients differ, even after taking into account the appropriate
constant rescalings of fields and parameters due to convention
differences.


\section{The Map Between $SL(4,R)/SO(4)$ and $SO(3,3)/SO(3)\times
SO(3)$}


Let us denote the $SL(4,R)/SO(4)$ coset representative by $\cV_\a^R$
which is a $4\times 4$ unimodular real matrix with inverse
$\cV^\a_R$:
 \be
 \cV^\a_R\cV_\a^S=\delta_R^S\ ,\qquad \a=1,...4,\quad R=1,...,4\ .
 \ee

The map between $\cV_\a^R$ and the $SO(3,3)/SO(3)\times SO(3)$ coset
representative $L_I^A$ can be written as
 \be
 L_I^A=\ft14\,\C_I^{\a\b}\,\eta^A_{RS}\,\cV_\a^R\,\cV_\b^S
 \nn\w2
 \equiv \ft14\,\cV\C_I\eta^A\cV\ ,
 \ee

where $\C^I$ and $\eta^A$ are the chirally projected  $SO(3,3)$
Dirac matrices which satisfy \cite{Howe:1983fr}
 \be
 (\C^I)_{\a\b} (\C^J)^{\a\b}=-4\eta^{IJ}\ ,\qquad
 (\C^I)_{\a\b}(\C_I)_{\c\d}=-2\e_{\a\b\c\d}\ ,
 \ee

where $\eta_{IJ}$ as well as $\eta_{AB}$ have signature $(---+++)$.
Similar identities are satisfied by $(\eta^A)_{RS}$. Both $\C^I$ and
$\eta^A$ are antisymmetric. Pairs of antisymmetric indices are
raised and lowered by the $\e$ tensor:
 \be
 V^{\a\b}=\ft12\e^{\a\b\c\d}V_{\c\d}\ ,\qquad V_{\a\b}=\ft12\e_{\a\b\c}V^{\c\d}\ .
 \ee

Since $\cV$ is real, the $\C$ and $\eta$-matrices must be real as
well. A convenient such representation is given by
 \be
 \C^I \equiv (\C^I)_{\a\b}= \pmatrix{\alpha^i\cr \beta^r\cr}\ ,\quad
 (\C^I)^{\a\b}= \pmatrix{\alpha^i\cr -\beta^r\cr}\ ,
\ee

where $\alpha^i$ and $\beta^r$ are real antisymmetric $4\times 4$
matrices that satisfy
 \bea
 && \alpha_i\alpha_j=\e_{ijk}~\alpha_k -\delta_{ij}\,\oneone\ ,\qquad
 (\alpha^i)_{\a\b}=\ft12 \e_{\a\b\c\d}\,(\alpha^i)_{\c\d}\ ,
 \label{ids}\w2
 && \beta_r\beta_s=\e_{rst}~\beta_t -\delta_{rs}\,\oneone\ ,\qquad
 (\beta^r)_{\a\b}= - \ft12 \e_{\a\b\c\d}\,(\beta^r)_{\c\d}\ .\nn
 \eea

Further useful identities are
 \bea
 &&
 (\a^i)_{\a\b}~(\a^i)_{\c\d}=\delta_{\a\c}\d_{\b\d}-\delta_{\a\d}\d_{\b\c}
 +\e_{\a\b\c\d}\ ,\la{ab1}
 \w2
 && \e^{ijk}(\a^j)_{\a\b}~(\a^k)_{\c\d}=\d_{\b\c}~(\a^i)_{\a\d}+{\rm 3\
 more}\ ,
 \label{ab2}
 \w2
 &&(\b^r)_{\a\b}~(\b^r)_{\c\d}=\delta_{\a\c}\d_{\b\d}-\delta_{\a\d}\d_{\b\c}
 -\e_{\a\b\c\d}\ ,\label{ab3}
 \w2
 && \e^{trs}(\b^r)_{\a\b}~(\b^s)_{\c\d}=\d_{\b\c}~(\b^t)_{\a\d}+{\rm 3\
 more}\ .\label{ab4}
 \eea

Using the above relations and recalling that $\cV$ is unimodular, it
simple to verify that
 \be
 L_I^IL_J^B\eta_{AB}=\eta_{IJ}\ ,\quad\quad L_I^A
 L_J^B\eta^{IJ}=\eta^{AB}\ .
 \ee

As a further check, let us compare the potential
 \be
 V=\ft14 e^{-\sigma}\left(C^{ir}C_{ir}-\ft19 C^2\right)
 \label{pot}
 \ee

for the $SO(4)$ gauged theory with that of \cite{Salam:1983fa} where
it is represented in terms of the $SL(4,R)$ coset representative. To
begin with, the function $C$ can be written as
 \bea
 C &=&-\ft1{\sqrt 2} f_{IJ}{}^K L^I_i L^J_j L_{K k}\,\epsilon^{ijk}\ ,
 \nn\w2
 &=& -\ft1{64\sqrt 2} f_{IJK} \left(\cV\C^I \eta_i\cV \right)
 \left( \cV\C^J\eta_j\cV\right)\left(\cV\C^K\eta_k\cV\right)\,\e^{ijk}
 \nn\w2
 &=& \ft1{8\sqrt 2} f_{IJK}\left[(\C^{IJK})_{\a\b}\,T^{\a\b}
 +(\C^{IJK})^{\a\b}\,T_{\a\b}\right]\ ,
 \eea

where
 \be
 T_{\a\b}=\cV_\a^R\,\cV_\b^S\,\d_{RS}\ ,\qquad
 T^{\a\b}=\cV^\a_R\,\cV^\b_S\,\d_{RS}\ .
 \label{b10}
  \ee

In the last step we have used \eq{ab2}.
In fact, the expression \eq{c1} is valid for any gauging, not
withstanding the fact that the $SO(4)$ invariant tensor $\delta_{RS}$
occurs in \eq{b10}. However, only for $SO(4)$ gauging in which
the $f_{IJK}$ refers to the $SO(4)$ structure constants,
\eq{c1} simplifies to give a direct relation between $C$ and
$T=T^{\a\b}\delta_{\a\b}$ that is manifestly $SO(4)$ invariant,
as will be shown below. To obtain a similar
relation for gaugings other than $SO(4)$, for example $SO(2,2)$, we would
need to construct the $\C$ and $\eta$ matrices in
a $SO(2,1)\times SO(2,1)$ basis with suitable changes in \eq{ids}. In that case,
the $SO(2,2)$ invariant tensor $\eta_{RS}$ would replace  the $SO(4)$
invariant tensor $\delta_{RS}$ in \eq{b10} and we could get a manifestly
$SO(2,2)$ invariant direct relation between $C$ and $T$.

In the case of $SO(4)$ gauging we have
$f_{IJK}=(\e_{ijk},-\e_{rst})$. Using this in \eq{c1} we find that
the $\e_{ijk}$ term gives a contribution  of the form $(\d_{\a\b}
T^{\a\b}+\d^{\a\b}T_{\a\b})$, while the $\e_{rst}$ term gives a
contribution of the form $(\d_{\a\b} T^{\a\b}-\d^{\a\b}T_{\a\b}$).
The $\d^{\a\b}T_{\a\b}$ contributions cancel and we are left with
 \be
 C=-\ft3{2\sqrt 2}\, T\ ,\qquad T\equiv T^{\a\b}\,\d_{\a\b}\ .
 \label{cf1}
 \ee

Similarly, it follows from the definition of $C^{ir}$ and the
orthogonality relations satisfied by $L_I^A$ that
 \bea
 C^{ir}C_{ir}&=& f_{IJK}f_{MN}{}^K\,L^I_iL^J_jL^M_iL^N_j+\ft13C^2\ .
 \label{cf2}
\eea

Thus, it suffices to compute
 \bea
 f_{IJK}f_{MN}{}^K\,L^I_iL^J_jL^M_iL^N_j &=&
 -\ft14\left(\cV\C^i\eta^j\cV\right)
 \left(\cV\C_i\eta_j\cV\right)+6
 \nn\w2
 &=& \ft12 T_{RS}T^{RS}-\ft12T^2\ .
 \label{cf3}
 \eea

 Using the results \eq{cf1}, \eq{cf2} and \eq{cf3} in \eq{pot}, we
 find
 \be
 V=\ft18 e^{-\sigma}\left(T_{RS}T^{RS}-\ft12 T^2\right)\ ,
 \ee

which agrees with the result of \cite{Salam:1983fa}.

In the case of $Sp(1)_R$ gauged $6D$ supergravity obtained from the
$SO(3,1)$ gauged supergravity in $7D$, i.e.~model II in section 3.2,
we have $f_{IJK}=(-\e_{rst}, -\e_{ijr})$, where $\e_{ijr}$ is
totally antisymmetric and $\e_{124}=\e_{235}=\e_{316}=1$. For this
case, the $C$-function has a more complicated form in terms of the
$SL(4,R)$ coset representative $\cV$. However, setting the scalar
fields equal to zero, which is required for model II at hand, $\cV$
becomes a unit matrix and the $C$-function vanishes. This is easily
seen in the first line of \eq{c1}, while it can be seen from the
last line of \eq{c1} by noting that the $\e_{rst}$ term gives the
contribution $(\d_{\a\b} T^{\a\b}-\d^{\a\b}T_{\a\b})$, and the
$\e_{rsi}$ term give the structure $({\vec \a}\cdot {\vec
\b})_{\a\b}T^{\a\b}+({\vec \a}\cdot {\vec \b})^{\a\b}T_{\a\b}$,
where ${\vec \b}$ refers to $\b_{r-3}$, both of which vanish when
$\cV$ is taken to be a unit matrix. In the second term this is due
to the fact that ${\vec \a}\cdot {\vec \b}$ is traceless.


\section{Dualization of the Axions in the Ungauged $6D$ Model}


The $(p+3)$ axionic scalars occurring in the $6D$ Lagrangian ${\cal
L}={\cal L}_B+{\cal L}_F$ with ${\cal L}_B$ and ${\cal L}_F$ given
in \eq{b} and \eq{f} can be dualized to $4$-form potentials with
tensor gauge symmetry straightforwardly. Start by adding the
suitable total derivative term to this Lagrangian to define
 \be
 {\mathcal L}_4={\mathcal L}_B+{\mathcal L}_F + \ft1{5!\sqrt
 6}\,\e^{\mu_1\cdots\mu_6}  \left(-H^i_{\mu_1\cdots \mu_5} {\cP}^i_{\mu_6}+
 H^r_{\mu_1\cdots \mu_5} {\cP}^r_{\mu_6}\right)e^{-\varphi}\ ,
 \label{L4}
 \ee

where the definitions \eq{ax} are to be used without the gauge
coupling constants. Recalling that \eq{br} holds, the $\Phi^I$ field
equation implies $dH^I_5=0$ with $H_5^i=H_5^IL_I^i$ and
$H_5^r=H_5^IL_I^r$, which means that locally
 \be
 H^I_{\mu_1\cdots\mu_5}= 5\partial_{[\mu_1} C^I_{\mu_2\cdots\mu_5]}\
 ,\qquad I=1,...,p+3\ .
 \ee

Solving for $(\cP_\mu^i, P_\mu^r)$ gives
 \bea
 \cP_\mu^i &=& \ft{\sqrt 2}{5!} \,\e^{\mu\nu_1\cdots\nu_5}
 H^i_{\nu_1\cdots\nu_5} - {\bar\psi}\c^\nu\c_\mu\sigma^i\psi_\nu
 -\ft12 X_\mu^i\ ,
 \nn\w2
 \cP_\mu^r &=& -\ft{\sqrt 2}{5!}\,\e^{\mu\nu_1\cdots\nu_5}
 H^i_{\nu_1\cdots\nu_5} +i {\bar\psi}^r\c^\nu\c_\mu\sigma^i\psi_\nu
 -2i X_\mu^r\ .
 \label{de2}
 \eea

Substituting these back into the Lagrangian \eq{L4}, we get
 \bea
 {\cal L}_4&=&{\cal L'}-\ft1{2\times 5!} e^{-2\varphi}
 H^i_{\mu_1\cdots\mu_5}H^{i\mu_1\cdots\mu_5}-\ft1{2\times 5!} e^{-2\varphi}
 H^r_{\mu_1\cdots\mu_5}H^{r\mu_1\cdots\mu_5}
 \nn\w2
 && -\ft1{5!\sqrt 2} e^{-\varphi} \e^{\mu\nu_1\cdots\nu_5}
 H^i_{\nu_1\cdots\nu_5}
 \left( {\bar\psi}\c^\nu\c_\mu\sigma^i\psi_\nu +\ft12 X_\mu^i\right)
 \nn\w2
 &&
 -\ft1{5!\sqrt 2} e^{-\varphi}\e^{\mu\nu_1\cdots\nu_5} H^r_{\nu_1\cdots\nu_5}
 \left( i {\bar\psi}^r\c^\nu\c_\mu\sigma^i\psi_\nu -2i X_\mu^r \right)\ ,
 \eea

where ${\cal L'}$ is the $(\cP_\mu^i, P_\mu^r)$ independent part of
${\mathcal L}={\mathcal L}_B+{\cal L}_F$ with ${\mathcal L}_B$ and
${\mathcal L}_F$ given in \eq{b} and \eq{f}. Thus, we have
${\mathcal L}_4={\mathcal L}_{4B}+{\mathcal L}_{4F}$ with\\
 \bea
 %
 %
 e^{-1}{\mathcal L}_{4B}&=& \ft{1}{4}R - \frac14 (\p _\mu \sigma)^2
          - \ft{1}{12} e^{2\sigma}
          G_{\mu\nu\rho} G^{\mu\nu\rho}
          - \ft{1}{8} e^{\sigma} F_{\mu\nu}^{r'} F^{\mu\nu r'}
   \label{bL}\w2
          &&
          -\ft14 \p_\mu\varphi \p^\mu\varphi -\ft18 P_\mu^{ir}P^\mu_{ir}
       - \ft1{2\times 5!} e^{-2\varphi}a_{IJ}
         H^I_{\mu_1\cdots\mu_5}H^{J\mu_1\cdots\mu_5}\ ,
  \nn\w4
%
%
e^{-1} \mathcal{L}_F &=&
          - \ft{i}{2} {\bar\psi}_\mu
          \c^{\mu\nu\rho}D_\nu\psi_\rho
          - \ft{i}2 {\bar\chi} \c^{\mu} D_\mu \chi
          - \ft{i}{2}{\bar\lambda}^{r'}\c^{\mu} D_\mu\lambda_{r'}
    \nn\w2
          &&- \ft{i}{2} {\bar\psi} \c^\mu D_\mu\psi
          - \ft{i}{2}{\bar\psi}^r\c^\mu D_\mu
          \psi^r - \ft{i}2{\bar\chi}\c^\mu\c^\nu\psi_\mu \p_\nu\sigma
    \nn\w2
          && -\ft12 {\bar\psi}^r\c^\mu\c^\nu\sigma_i\psi_\mu P_\nu^{ir}
          +\ft{i}2{\bar\psi}\c^\mu\c^\nu\psi_\mu
          \p_\nu\varphi
          \nn\w2
          &&
          -\ft1{5!\sqrt 2} e^{-\varphi}
          H^I_{\mu_1\cdots\mu_5}
          \left(
          {\bar\psi}\c^\nu\c_{\mu_1\cdots\mu_5}\sigma^i\psi_\nu\,L_I^i
          + {\bar\psi}^r\c^\nu\c_{\mu_1\cdots\mu_5}
          \sigma^i\psi_\nu\,L_I^r\right)
          \nn\w2
          && +\ft{i}{24} e^{\sigma}G_{\mu\nu\rho} X^{\mu\nu\rho}
          - \ft{i}{4} \eps  F^{r'}_{\mu\nu} X^{\mu\nu}_{r'}
          +\ft1{2\sqrt 2\times 5!} e^{-\varphi} H^I_{\mu_1\cdots\mu_5}
          X_I^{\mu_1\cdots\mu_5}\ ,
\label{fL}
  \eea

where the structure constants (hence the $C$-functions as well) are
to be set to zero in the definitions \eq{mc6} and \eq{cd6}, and\\
 \bea
 X^{\mu\nu\rho} &=&
      {\bar\psi}^\lambda \c_{[\lambda}\c^{\mu\nu\rho}
      \c_{\tau]}\psi^\tau
      + 2{\bar\psi}_\lambda\c^{\mu\nu\rho}
      \c^\lambda \chi
      - 2{\bar\chi}\c^{\mu\nu\rho}\chi
      + {\bar\lambda}^{r'}\c^{\mu\nu\rho}\lambda_{r'}
      + {\bar\psi}^r\c^{\mu\nu\rho}\psi_r
      +  {\bar\psi}\c^{\mu\nu\rho}\psi\ ,
      \nn\w2
X^{\mu\nu}_{r'} &=&
      {\bar\psi}_\rho\c^{\mu\nu}\c^\rho\lambda_{r'}
      +  {\bar\chi}\c^{\mu\nu}\lambda_{r'}\ ,
      \nn\w2
X^{\mu_1\cdots\mu_5}_I &=&
      L_I^i\, \big( {\bar\psi}^\rho \c_{[\rho}\c^{\mu_1\cdots\mu_5}\c_{\tau]}
      \sigma_i\psi^\tau + 2 {\bar\chi} \c^{\mu_1\cdots\mu_5}\sigma_i \chi
      + {\bar\lambda}^{r'} \c^{\mu_1\cdots\mu_5}\sigma_i\lambda_{r'}
      \nn\w2
      &&
      - {\bar\psi}^r \c^{\mu_1\cdots\mu_5}\sigma_i\psi_r
      -{\bar\psi}\c^{\mu_1\cdots\mu_5}\sigma_i\psi\big)
      -4i L_I^r \,{\bar\psi}\c^{\mu_1\cdots\mu_5}\psi_r \ .
 \label{xdefs2}
 \eea

\bigskip

The action is invariant under the following supersymmetry
transformations:
 \bea
 %
 %
 \delta e_{\mu}^{m} &=&
 i\bar{\epsilon}\c^{m}\psi{}_{\mu}\ ,
 \nn\w2
 \delta \psi_{\mu} &=& D_{\mu} \e -\ft{1}{24}e^{\sigma}
 \c^{\rho\sigma\tau}\c_{\mu}G_{\rho\sigma\tau}\,\e
 -\ft{i}{5!\sqrt 6} e^{-\varphi} \c^{\mu\nu_1\cdots\nu_5} H^i_{\nu_1\cdots\nu_5}\e \ ,
 \nn\w2
 \delta \chi &=&
 - \ft12 \c^{\mu} \p_{\mu} \sigma\e
 - \ft{1}{12} e^{\sigma}
 \c^{\rho\sigma\tau} G_{\rho\sigma\tau}\,\e \ ,
 \nn\w2
 \delta B_{\mu\nu}&=&
 ie^{-\sigma}\left(\bar{\epsilon}\c_{[\mu}\psi_{\nu]}
 + \ft12 \bar{\epsilon}\c_{\mu\nu} {\chi}\right)
 - A_{[\mu}^{r'} \delta A_{\nu]}^{r'} , \nn\w2
 \delta \sigma &=& -i\bar{\epsilon} \chi\ ,
 \nn\w4
 %
 %
 \delta A_{\mu}^{r'} &=& i e^{- \sigma/2}
 \bar{\epsilon}\c_{\mu} \lambda^{r'}\ ,
 \nn\w2
 \delta \lambda^{r'} &=&
 - \ft{1}{4} \eps \c^{\mu\nu}F_{\mu\nu}^{r'}\e\ ,
 \nn\w2
 %
 %
 \delta C^I_{\mu_1\cdots\mu_4}&=& -\ft1{\sqrt 2}\left(
 {\bar\e}\c_{\mu_1\cdot\mu_4}\sigma^i\psi
 +4{\bar\e}\c_{[\mu_1\cdot\mu_3}\sigma^i\psi_{\mu_4]} \right)L^I_i
 -\ft{i}{\sqrt 2}{\bar\e}\c_{\mu_1\cdot\mu_4}\psi^r L^I_r\ ,
 \nn\w2
 %
 %
 L^I_i\delta L_I^r &=& -\bar{\epsilon} \sigma_i\psi^r\ ,
 \nn\w2
 \delta \varphi &=& i \bar{\epsilon}\psi,
 \nn\w2
 \delta {\psi}&=&
 \ft12 \c^\mu \p_\mu\varphi \epsilon
 +\ft{i}{5!\sqrt 2} e^{-\varphi} \c^{\mu_1\cdots\mu_5}
 H^i_{\mu_1\cdots\mu_5}\e\ ,
 \nn\w2
 \delta \psi^r &=&
 \ft{i}2 \c^\mu P_\mu^{ir}  \sigma_i
 -\ft{i}{5!\sqrt 2} e^{-\varphi} \c^{\mu_1\cdots\mu_5}
 H^r_{\mu_1\cdots\mu_5} \e\ .
 \label{susy3}
 \eea

The supersymmetry transformation rule for $C^I_{\mu_1\cdots\mu_4}$
is derived from the requirement of supercovariance of \eq{de2},
which requires the cancellation of the $\partial_\mu\e$ terms.


\section{The Iwasawa Decomposition of $SO(p,q)$}


We begin with the Iwasawa decomposition of the $SO(n+3,3)$ algebra
as $g=h \oplus {\it a} \oplus {\it n}$ where
 \bea
 &&h:\qquad X_{ij}\ , Y_{ij}\ , Z_{ir}\ , T_{rs}\ ,
 \nn\w2
 && {\it a}:\qquad H_i\ ,
 \nn\w2
 && {\it n}:\qquad E_i{}^j\ , V^{ij}\ , U_{ir}\ ,\quad i>j\ .
 \eea

Here $X=E-E^T$, $Y=V-V^T$, $Z=U-U^T$, together with the $SO(n)$
generators $T_{rs}=-T_{sr}$ form the maximal compact subalgebra
$\{h\}$ of $SO(n+3)\times SO(3)$. Furthermore, $\{a\}$ are the
noncompact Cartan generators and $\{n\}$ are the remaining
noncompact generators of $SO(n+3,3)$. The generators $a\oplus n$
form the solvable subalgebra of $SO(n+3,3)$, and can be represented
as (see, for example, \cite{Lu:1998xt})
 \bea
 &&\vec H = \left(\begin{array}{c|c|c}
       \sum_i \vec c_i\, e_{ii} & 0 & 0 \\ \hline
        0 & 0 & 0 \\ \hline
        0 & 0 & -\sum_i \vec c_i\, e_{ii}
        \end{array}\right) \ ,\qquad
 E_i{}^j = \left(\begin{array}{c|c|c}
                 -e_{ji} & 0 & 0\\ \hline
                        0 & 0 & 0\\ \hline
                        0 & 0 & e_{ij}
         \end{array}\right)\ ,\nn\\
 &&V^{ij} = \left(\begin{array}{c|c|c}
                  0 & 0 &  e_{ij}- e_{ji} \\ \hline
                        0 & 0 & 0 \\ \hline
                        0 & 0 & 0
         \end{array}\right)\ ,\qquad
 U^i_r = \left(\begin{array}{c|c|c}
                  0 & e_{ir} &  0 \\ \hline
                        0 & 0 & e_{r i} \\ \hline
                        0 & 0 & 0
         \end{array}\right)\ .
 \label{generators}
 \eea

 The maximal compact subalgebra generators are then represented as
 \bea
 &&X_{ij}= \left(\begin{array}{c|c|c}
                 e_{ij}-e_{ji} & 0 & 0\\ \hline
                        0 & 0 & 0\\ \hline
                        0 & 0 & e_{ij}-e_{ij}
         \end{array}\right)\ ,\qquad
 Y_{ij} = \left(\begin{array}{c|c|c}
                  0 & 0 &  e_{ij}- e_{ji} \\ \hline
                        0 & 0 & 0 \\ \hline
                         e_{ij}- e_{ji}& 0 & 0
         \end{array}\right)\ ,
 \nn\w2
 &&Z_{ir} = \left(\begin{array}{c|c|c}
                  0 & e_{ir} &  0 \\ \hline
                        -e_{ri}& 0 & e_{r i} \\ \hline
                        0 & -e_{ir}& 0
         \end{array}\right)\ ,\qquad
         T_{rs} = \left(\begin{array}{c|c|c}
                  0 & 0 &  0 \\ \hline
                        0& e_{rs}-e_{sr} & 0 \\ \hline
                        0 & 0 & 0  \end{array}\right)\ .
 \eea

Each $e_{ab}$ is defined to be a matrix of the appropriate dimensions
that has zeros in all its entries except for a 1 in the entry at row $a$
and column $b$.  These satisfy the matrix product rule $e_{ab}\, e_{cd} =
\delta_{bc}\, e_{ad}$.

The solvable subalgebra of $SO(n+3,3)$ has the nonvanishing
commutators
 \bea
 && [{\vec H},E_i{}^j]={\vec b}_{ij}~E_i{}^j\ ,\qquad
    [{\vec H},V^{ij}]={\vec a}_{ij}~V^{ij}\ ,\qquad
    [{\vec H},U_r^j]={\vec c}_i~U_r^i\ ,
    \nn\w2
    &&[E_i{}^j,E_k{}^\ell]=\delta^j_k E_i{}^\ell -\delta_i^\ell
    E_k{}^j\ ,
    \nn\w2
    && [E_i{}^j,V^{k\ell}]=-\delta^k_i V^{j\ell}-\delta^\ell_i V^{kj}\
    ,
    \qquad
    [E_i{}^j,U_r^k]=-\delta_i^kU_r^j\ ,
    \nn\w2
    && [U_r^i,U_s^j]=\delta_{rs}V^{ij}\ ,
    \label{cr1}
 \eea

where the structure constants are given by
 \be
 {\vec b}_{ij}={\sqrt 2}~(-{\vec e}_i + {\vec e}_j)\ ,\qquad
 {\vec a}_{ij}={\sqrt 2}~( {\vec e}_i + {\vec e}_j)\ ,\qquad
 {\vec c}_i={\sqrt 2}~{\vec e}_i\ .
 \label{sc}
 \ee

The nonvanishing commutation commutation rules of the maximal
compact subalgebra $SO(n+3)\oplus SO(3)$ are
 \bea
 &&[X_{ij}, X_{k\ell}]=\d_{jk}X_{i\ell} + 3\ perms\ ,
   \quad\quad [T_{pq},T_{rs}]=\d_{qr}T_{ps}+ 3\ perms\ ,
 \w2
 && [X_{ij},Y_{k\ell}]=\d_{jk} Y_{i\ell} + 3\ perms\ ,
 \quad\quad [X_{ij}, Z_{kr}]=\d_{jk} Z_{ir}-\d_{ik} Z_{jr}\ ,
 \nn\w2
  && [Y_{ij}, Y_{k\ell}]=\d_{jk}X_{i\ell} + 3\ perms\ ,
     \quad\quad [T_{pq},Z_{ir}]=\d_{qr} Z_{ip}-\d_{pr}Z_{iq}\ ,
     \nn\w2
     && [Z_{ir}, Z_{js}]=-\d_{rs} X_{ij}+\d_{rs} Y_{ij}-2\d_{ij} T_{rs}\
     ,\qquad
     [Y_{ij}, Z_{kr}]=-\d_{jk}Z_{ir}+\d_{ik}Z_{jr}\ .
 \nn
 \eea

\end{appendix}

\newpage



\begin{thebibliography}{10}



\bibitem{Nishino:1984gk}
H.~Nishino and E.~Sezgin, ``Matter and gauge couplings of N=2
supergravity in six-dimensions'', Phys.\ Lett.\ B {\bf 144} (1984)
187;

H.~Nishino and E.~Sezgin, ``The complete N=2, D = 6 supergravity
with matter and Yang-Mills couplings'', Nucl.\ Phys.\ B {\bf 278}
(1986) 353.

\bibitem{Randjbar-Daemi:1985wc}
S.~Randjbar-Daemi, A.~Salam, E.~Sezgin and J.~Strathdee, ``An
anomaly free model in six-dimensions'',  Phys.\ Lett.\ B {\bf 151}
(1985) 351.

\bibitem{Salam:1985mi}
A.~Salam and E.~Sezgin, ``Anomaly freedom in chiral
supergravities'', Phys.\ Scripta {\bf 32} (1985) 283.

\bibitem{Bergshoeff:1986hv}
E.~Bergshoeff, T.~W.~Kephart, A.~Salam and E.~Sezgin, ``Global
anomalies in six-dimensions'', Mod.\ Phys.\ Lett.\ A {\bf 1} (1986)
267.

\bibitem{Avramis:2005qt}
S.~D.~Avramis, A.~Kehagias and S.~Randjbar-Daemi, ``A new
anomaly-free gauged supergravity in six dimensions'',
arXiv:hep-th/0504033.

\bibitem{Avramis:2005hc}
S.~D.~Avramis and A.~Kehagias,
``A systematic search for anomaly-free supergravities in six dimensions'',
arXiv:hep-th/0508172.


\bibitem{Cvetic:2003xr}
M.~Cvetic, G.~W.~Gibbons and C.~N.~Pope, ``A string and M-theory
origin for the Salam-Sezgin model'', Nucl.\ Phys.\ B {\bf 677}
(2004) 164 [arXiv:hep-th/0308026].

\bibitem{Hull:1984vg}
  C.~M.~Hull,
  ``Noncompact Gaugings of N=8 supergravity'',
  Phys.\ Lett.\ B {\bf 142} (1984) 39.


\bibitem{Hull:1988jw}
  C.~M.~Hull and N.~P.~Warner,
  ``Noncompact gaugings from higher dimensions,''
  Class.\ Quant.\ Grav.\  {\bf 5} (1988) 1517.


\bibitem{Salam:1983fa}
A.~Salam and E.~Sezgin, ``SO(4) Gauging of N=2 Supergravity in
seven-dimensions'', Phys.\ Lett.\ B {\bf 126} (1983) 295.


\bibitem{Cvetic:1999pu}
  M.~Cvetic, J.~T.~Liu, H.~Lu and C.~N.~Pope,
  ``Domain-wall supergravities from sphere reduction'',
  Nucl.\ Phys.\ B {\bf 560} (1999) 230 [arXiv:hep-th/9905096].


\bibitem{Lu:1999bc}
  H.~Lu and C.~N.~Pope,
  ``Exact embedding of N = 1, D = 7 gauged supergravity in D = 11'',
  Phys.\ Lett.\ B {\bf 467} (1999) 67 [arXiv:hep-th/9906168].


\bibitem{Bergshoeff:1985mr}
E.~Bergshoeff, I.~G.~Koh and E.~Sezgin, ``Yang-Mills / Einstein
supergravity in seven-dimensions,'' Phys.\ Rev.\ D {\bf 32} (1985)
1353.

\bibitem{Anguelova:2002kd}
  L.~Anguelova, M.~Rocek and S.~Vandoren,
  ``Hyperkaehler cones and orthogonal Wolf spaces'',
  JHEP {\bf 0205} (2002) 064 [arXiv:hep-th/0202149].

\bibitem{Theis:2003jj}
  U.~Theis and S.~Vandoren,
  ``N = 2 supersymmetric scalar-tensor couplings'',
  JHEP {\bf 0304} (2003) 042 [arXiv:hep-th/0303048].

\bibitem{Dall'Agata:2003yr}
  G.~Dall'Agata, R.~D'Auria, L.~Sommovigo and S.~Vaula,
  ``D = 4, N = 2 gauged supergravity in the presence of tensor
  multiplets'',
  Nucl.\ Phys.\ B {\bf 682} (2004) 243 [arXiv:hep-th/0312210].

\bibitem{Howe:1985sb}
  P.~S.~Howe, A.~Karlhede, U.~Lindstrom and M.~Rocek,
  ``The geometry of duality'', Phys.\ Lett.\ B {\bf 168} (1986) 89.


\bibitem{Park:1988id}
Y.~J.~Park, ``Gauged Yang-Mills-Einstein supergravity with three
index field in seven-dimensions'', Phys.\ Rev.\ D {\bf 38} (1988)
1087.

\bibitem{deRoo:1985jh}
  M.~de Roo and P.~Wagemans,
  ``Gauge matter coupling in N=4 supergravity'',
  Nucl.\ Phys.\ B {\bf 262} (1985) 644.

\bibitem{Lu:1998xt}
H.~Lu, C.~N.~Pope and K.~S.~Stelle, ``M-theory/heterotic duality: A
Kaluza-Klein perspective'', Nucl.\ Phys.\ B {\bf 548} (1999) 87
[arXiv:hep-th/9810159].

\bibitem{Cremmer:1997ct}E.~Cremmer, B.~Julia, H.~Lu and C.~N.~Pope,
``Dualisation of dualities. I'', Nucl.\ Phys.\ B {\bf 523} (1998) 73
[arXiv:hep-th/9710119].


\bibitem{Avramis:2004cn}
S.~D.~Avramis and A.~Kehagias, ``Gauged D = 7 supergravity on the
S(1)/Z(2) orbifold'', arXiv:hep-th/0407221.


\bibitem{Townsend:1983kk}
P.~K.~Townsend and P.~van Nieuwenhuizen, ``Gauged seven-dimensional
supergravity'',  Phys.\ Lett.\ B {\bf 125} (1983) 41.


\bibitem{Howe:1983fr}
P.~S.~Howe, G.~Sierra and P.~K.~Townsend, ``Supersymmetry in
six-dimensions'', Nucl.\ Phys.\ B {\bf 221} (1983) 331.





\end{thebibliography}
\end{document}